\documentclass{aa} 
\usepackage{graphicx}
\usepackage{natbib}
\bibpunct{(}{)}{;}{a}{}{,} 

\def\nh{N$_{\rm H}$}
\def\ergscm{ergs s$^{-1}$ cm$^{-2}$}
\def\ergs{ergs s$^{-1}$}
\def\xmm{XMM-Newton}
\def\omecen{$\omega$ Cen}
\def\pow{PL}
\def\brem{TB}
\def\bb{BB}
\def\nodata{...}

\begin{document}

\title{An XMM-Newton observation of the \\
       globular cluster Omega Centauri}

\author{Bruce Gendre
	\and Didier Barret
	\and Natalie A. Webb}
\offprints{Bruce Gendre, \email{gendre@cesr.fr}}
\institute{Centre d'Etude Spatiale des Rayonnements, 9 avenue
       du Colonel Roche, 31028 Toulouse CEDEX 4, France}

\date{Received $<$date$>$ / Accepted $<$date$>$}

\abstract{

We report on a deep XMM-Newton EPIC observation of the globular cluster Omega Cen performed on August 13th, 2001. We have detected 11 and 27 faint X-ray sources in the core and half mass radii, searching down to a luminosity of 1.3 $\times 10^{31}$ \ergs\ in the 0.5-5 keV range. Most sources have bolometric X-ray luminosities between $\sim 10^{31}-10^{32}$\ergs. We present the color-color and hardness-intensity diagrams of the source sample, as well as high-quality EPIC spectra of the brightest objects of the field; including the two candidate Cataclysmic Variables (CVs) in the core and the quiescent neutron star low-mass X-ray binary candidate. The spectra of the latter objects fully support their previous classification. We  show that the bulk of sources are hard and spectrally similar to CVs. The lack of soft faint sources might be related to the absence of millisecond pulsars in the cluster. The  XMM-Newton observations reveal the presence of an excess of sources well outside the core of the cluster where several RS CVn binaries have already been found.  We have also analyzed a publicly available Chandra ACIS-I observation performed on January 24-25th, 2000, to improve the \xmm\ source positions and to search for source intensity variations between the two data sets.  63 \xmm\ sources have a Chandra counterpart, and 15 sources within the half-mass radius have shown time variability. Overall, the general properties of the faint X-ray sources in \omecen\ suggest that they are predominantly CVs and active binaries (RS CVn or BY Dra).


}

\titlerunning{An XMM-Newton observation of Omega Centauri}
\authorrunning{Gendre et al.}
\maketitle

\section{Introduction}
Omega Centauri (\object{NGC 5139}, \omecen) is one of the best studied objects of our
galaxy. It is the most massive globular cluster \citep[5.1 $\times 10^{6}$
M$_{\sun}$, ][]{mey95}. It is characterized by large core and half mass
radii \citep[154.88 $\arcsec$ and 250.8 $\arcsec$ respectively,][]{har96}.
Binaries are expected to be present in \omecen\, either as a result of the
evolution of primordial binaries, or through close encounters between stars
in the cluster \citep{dis92,dis94,dav95,ver02}. Binaries such as CVs, 
low mass X-ray binaries either
with a neutron star or a black hole, or active X-ray binaries
(RS CVn or BY Dra systems), millisecond pulsars could thus form. Some
of these binaries have already been found as faint X-ray sources in \omecen\ \citep{ver01}.

Faint X-ray sources were first detected in \omecen\ by the EINSTEIN X-ray satellite. EINSTEIN detected 5 faint point sources \citep[one in the core,][]{her83} and a possible extended emission region, within the half mass radius \citep{har82}.  A decade after EINSTEIN, ROSAT detected 22 faint sources in the line of sight of $\omega$ Cen \citep{joh94,ver00}. ROSAT confirmed the EINSTEIN sources, and resolved the core source into three components \citep{ver00}. However, ROSAT did not find any evidence for the diffuse emission seen by EINSTEIN \citep{joh94}.

More recently, Chandra observed $\omega$ Cen and detected over 140 faint X-ray sources \citep{coo02}. From follow-up observations using the accurate Chandra positions, \citet{coo02} claimed that there were at least three classes of binaries present in the detected sample. Two of the three ROSAT core sources (ROSAT R9a and R9b) may be CVs \citep{car00}. The third ROSAT core source \citep[R20,][]{ver00} was associated with a main sequence optical counterpart, showing weak H$\alpha$ emission, suggesting a BY Dra system rather than a CV. In addition, two more Chandra core sources were detected with HST/WFPC2, with properties matching those of RS CVn or active-corona binaries \citep{coo02}. Finally, \citet{coo02} found the X-ray counterparts of two variable binaries discovered far out from the cluster center by \citet{kal96}. Based on their light curve properties, these two systems are proposed to be RS CVn binaries  \citep{kal96,kal02}. Beside these three classes of binaries, by looking at the spectral characteristics of the Chandra sources, \citet{rut02} noticed that one relatively bright object had an extremely soft X-ray spectrum. This spectrum was found to be consistent with those observed from field quiescent neutron star binaries \citep{rut02}. X-rays would then come from the neutron star surface maintained at a high temperature by episodic mass accretion from a binary companion. In total, up to four different types of binaries may have already been found in the cluster.
 
We have initiated a survey of nearby globular clusters with XMM-Newton \citep[M22,][ M13, NGC 6366, Gendre et al., in preparation and $\omega$ Cen]{web02a}. Because of the limited angular resolution of XMM-Newton, we have selected nearby clusters with large core radii. Taking advantage of the large collecting area of XMM-Newton ($\sim$ 6 times that of Chandra), we wish to obtain the best possible spectral and timing information for the widest possible sample of faint X-ray sources.

In this paper, we present the first results of our deep \xmm\ observation of \omecen. We describe the general properties of the population of faint X-ray sources detected in the cluster (section \ref{prop_gen}). Using the publicly
available Chandra  observation, we have correlated the \xmm\ and Chandra data (sections \ref{xmm_chandra} and \ref{diffus}) to improve the \xmm\ positions and search for intensity variations between the two data sets (section \ref{vari}). We also present the spectra of the brightest objects in the field, with the emphasis on those for which an identification already exists (section \ref{xmm_spec}). We briefly discuss the implications of our findings in section \ref{discu}.

\section{General properties of the faint X-ray sources in Omega Cen}
\label{prop_gen}

\subsection{The XMM-Newton observation}
We observed $\omega$ Cen on 2001 August 13 with the XMM-Newton EPIC MOS \citep{tur01} and PN \citep{str01} cameras, using a full frame window mode and a medium filter. The observation was 40 kilosecond long with a low and stable background. The data were analyzed with the latest version (5.3.3) of the XMM-Newton Science Analysis Software (SAS). We used the calibration chains of the EPIC cameras, {\it emchain} and {\it epchain}, using the {\it embadpixfind} task to flag bad pixels and bad columns. We filtered the event files produced for good time intervals and non astrophysical events (electronic noise, cosmic rays). We used the predefined patterns, keeping only patterns 0-12 for the MOS detectors and patterns 0-4 for the PN, and we rejected all the events flagged as 'bad' by the calibration chains. Finally, we also rejected events with energies below 0.4 keV and above 10 keV, because of a high number of bad patterns. 

\subsection{Source detection}
Sources were searched between 0.5 and 5.0 keV, a range which encloses the peak of the effective area of the EPIC cameras. A wavelet detection algorithm was used\footnote{see the {\it ewavelet} documentation  available at http://xmm.vilspa.esa.es/\-external/\-xmm\_user\_support/\-documentation/\-sas\_pkg\_frame.shtml}. It is better suited to crowded fields than the sliding box algorithm. Given the early development stage of the task, we used a conservative 4 $\sigma$ as the detection threshold. For each camera, the source list so obtained was used as an input to the task {\it emldetect}. {\it emldetect} computes for each source a maximum likelihood, taking into account the point spread function of the instrument. For each source, the task returns its best fit position, the statistical errors on this position ($1 \sigma$ or 68\% confidence level), its count rate, and a maximum likelihood detection value. We used a maximum likelihood threshold  of 12. In order to estimate the statistical error at the 90\% confidence level for the source positions, we modified the public version of {\it emldetect} following the recommendation of the task author (for a two parameter fit, the 90 \% confidence limit level is given by likelihood + 6.18, G. Lamer, private communication). Three PN sources were removed because their best fit positions fell onto bad columns. The cleaned PN, MOS1 and MOS2 source lists were then correlated using a customized version of the task {\it srcmatch}. {\it srcmatch} returns the positions of the correlated sources weighted by their statistical errors as derived for each instrument separately. The maximum likelihood of a correlated source is the sum of the individual maximum likelihoods.

\begin{figure}[!hb]
  \centerline{\includegraphics[width=8cm]{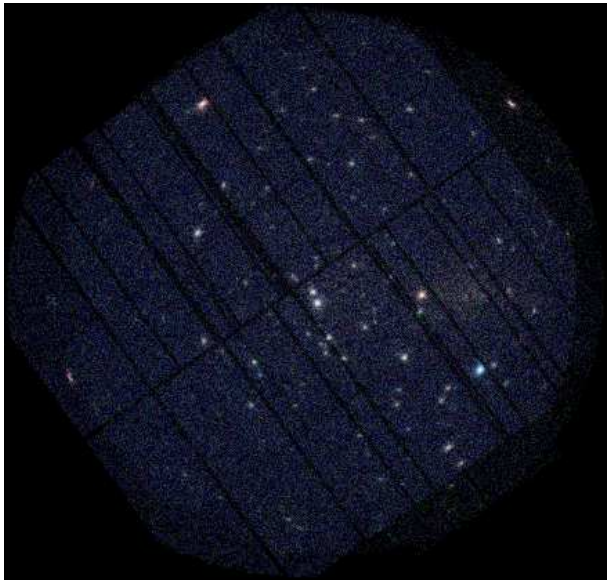}}
  \caption[]{A false color image of the XMM-Newton field of view. It combines the EPIC-PN and MOS images. The color bands we used were 0.5-1.5 keV (red), 1.5-3.0 keV (green) and 3.0-10.0 keV (blue).}
  \label{gendre_fig1}
\end{figure}

The majority of the correlated sources are detected by the PN camera because it is more sensitive than the MOS cameras. There are however some sources missed by the PN, due to CCD gaps, or bad columns and the smaller field of view of the PN compared to the MOS; these sources are marked MOS only in Tables \ref{gendre_table1} and \ref{gendre_table2}.

146 sources were detected by the EPIC cameras; 59 are seen only by the PN, 9 are MOS only (and all of them are seen in MOS 1 and MOS 2), and the remaining 78 are detected in the PN and at least one MOS camera. We present a false color combined PN and MOS image of the field of view in Fig. \ref{gendre_fig1}. The positions and statistical errors given at the 90\% confidence level, the 0.5-5.0 keV source count rate and associated error are listed in Table \ref{gendre_table1} for those sources lying within the half mass radius and in Table \ref{gendre_table2} for the remaining sources. In EPIC-PN, a count rate of $10^{-2}$ counts s$^{-1}$ corresponds to an unabsorbed flux of $\sim 2.7\times10^{-14}$ \ergscm\ for a 0.6 keV blackbody model absorbed through the interstellar absorption derived from the optical extinction \citep[\nh\ = $8.4\times 10^{20}$ cm$^{-2}$,][]{djo93,pre95}. Assuming a 3 keV thermal Bremsstrahlung and a power law of photon index 2, the corresponding fluxes are $\sim 2.5\times10^{-14}$ \ergscm\ and $\sim 2.4\times10^{-14}$ \ergscm\ respectively.

For the blackbody model, this flux translates to a luminosity of $\sim 10^{32}$ \ergs\ at the distance of \omecen\ \citep[5.3 kpc, ][]{har96,tho01}. Thus in Tables \ref{gendre_table1} and \ref{gendre_table2} one can see that the bulk of sources have luminosities in the range $10^{31}-10^{32}$ \ergs.

\subsection{Background sources}

Some of the sources listed in Tables \ref{gendre_table1} and \ref{gendre_table2} are extragalactic background sources unrelated to the cluster. In order to estimate their number, we used the statistical Log~N-Log~S relationship of extragalactic sources derived from the Lockman Hole \xmm\ data \citep[][]{has01}. To account for the vignetting function of the XMM-Newton mirrors, we have computed limiting count rates (for source detection) within different annuli (all centered on the cluster center), using an approach similar to \citet{coo02}. The radius of each annulus is computed such that the annulus contains a large ($\ge $ 25) number of sources (the radius varies between 2.5 and 4.2\arcmin). This allows us to set the limiting count rate to the count rate of the weakest source detected in that annulus. The limiting count rate is a factor of $\sim 2$ larger in the outer annulus than in the inner annulus. For direct comparison with the Log N - Log S curve of \citet{has01}, these count rates have then been converted into unabsorbed 0.5-2.0 keV fluxes using a power law model of index 2.0 absorbed through the cluster \nh.  Following this procedure, after a proper surface normalization, one expects 4, 9, 35 and 65 background sources within the core, half mass, twice the half mass and a 12.5\arcmin\ radii (the values so obtained were rounded to the nearest integer). Beyond 12.5\arcmin, where we do not expect any cluster sources, the number of detected sources matched the one estimated with this procedure. 

As an indication, we have computed the error on the above estimates assuming a 10\% uncertainty on the limiting count rate estimate and a 10\% uncertainty in the \xmm\ calibration (note that the Log~N-Log~S relationship was derived from a processing of the Lockman hole data with the SAS prior to its first public release). This gives an error of 1 and 2 on the estimated number of background sources within the core and half mass radii.

\subsection{Color-color and hardness-intensity diagrams}

\begin{figure*}[!t]
  \hspace*{-2cm}\includegraphics[width=21cm]{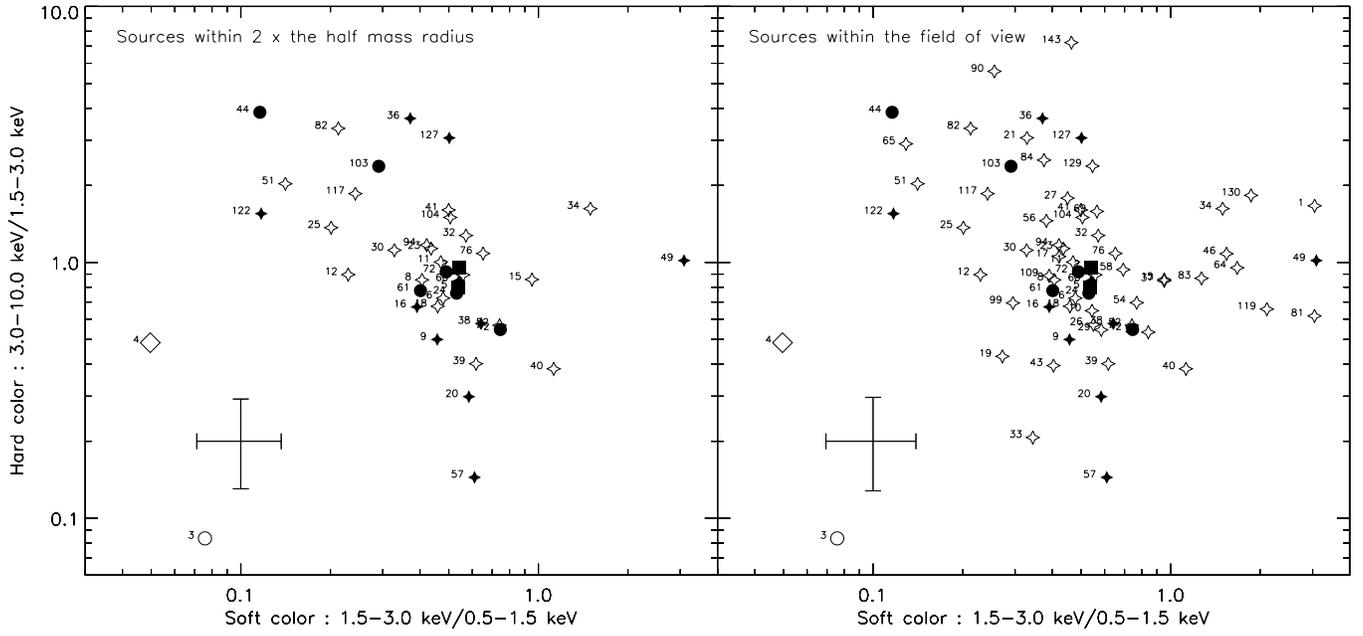}
  \caption[]{The color-color diagram of the sources detected by the EPIC-PN camera within twice the half mass radius (left) and within the field of view (right). The star identified by \citet{coo95}, the quiescent neutron star binary candidate and the two CV candidates are represented by an open circle, an open diamond and two filled squares respectively. Unknown sources are represented by a filled circle, a filled star and an open star if the source lies within the core radius, within the half mass radius, or outside the half mass radius respectively. A representative error bar is shown. Each source is labeled according to Tables \ref{gendre_table1} and \ref{gendre_table2}.}
  \label{gendre_fig2}
\end{figure*}

\begin{figure*}[!t]
  \hspace*{-2cm}\includegraphics[width=21cm]{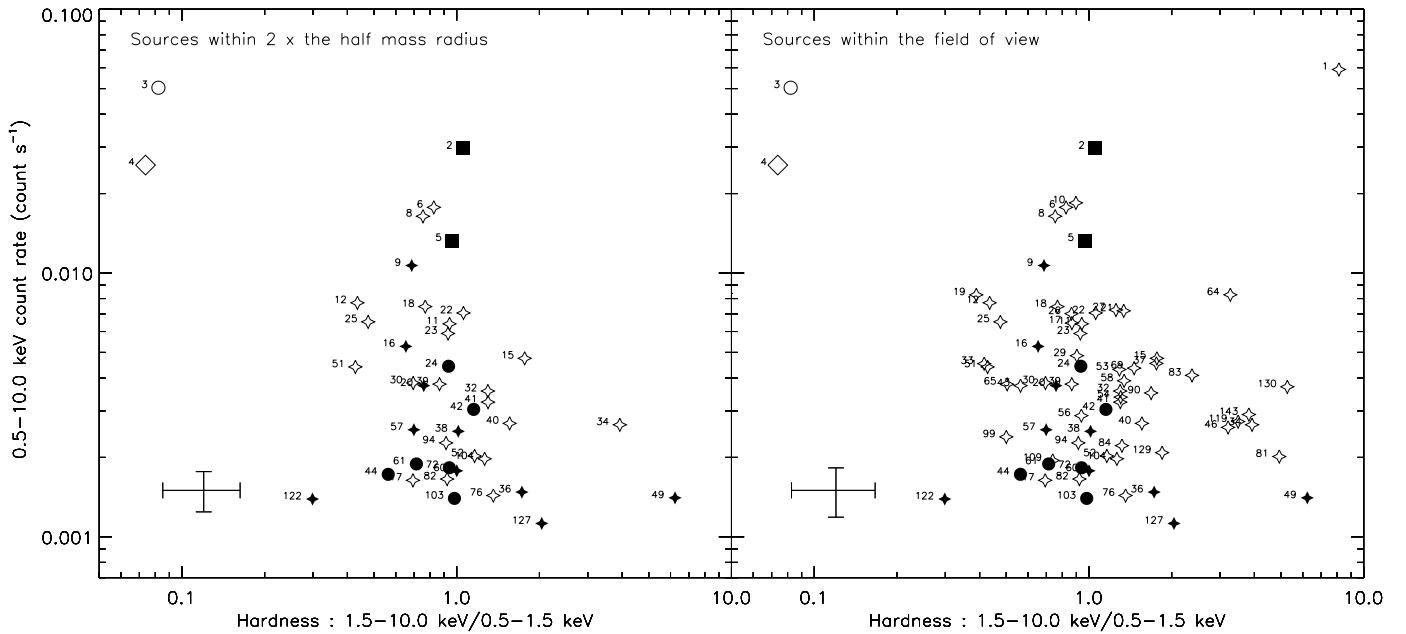}
  \caption[]{The hardness-intensity diagram of the sources detected by the EPIC-PN camera within twice the half mass radius (left) and within the whole field of view (right). The symbols refer to the same objects as Fig. \ref{gendre_fig2}. Each source is labeled according to Tables \ref{gendre_table1} and \ref{gendre_table2}. The intensity is corrected for the vignetting of the mirrors. A representative error bar (source 42) is shown.}
  \label{gendre_fig3}
\end{figure*}

In order to investigate the general properties of the sources detected by \xmm, we have computed X-ray color-color and hardness-intensity diagrams. For this purpose, we have produced PN images in three adjacent energy bands: 0.5-1.5 keV, 1.5-3.0 keV, 3.0-10.0 keV \citep[similar bands were used by][]{gri01a}. From these images, we have computed the net exposure corrected source count rate. To produce meaningful diagrams we have considered only sources detected with more than 3 counts in each band. There are 71 sources fitting this criteria. Two sets of diagrams have been computed: one for the sources found within a region of radius equal to twice the half mass radius and one for the whole field of view. They are presented in Fig. \ref{gendre_fig2} and \ref{gendre_fig3}.

\section{Cross-correlation with a previous Chandra observation}
\label{xmm_chandra}

The mean statistical error on the positions of the sources detected by \xmm\ is of the order of  $\sim 4$\arcsec\ (see Tables \ref{gendre_table1} and  \ref{gendre_table2}). To get the final positional error, one must add quadratically, the systematic error on the pointing direction of the XMM-Newton satellite, which is about 4\arcsec\ \citep{jan01}. This means that on average the position error will be around 6\arcsec.

\begin{figure*}[!t]
  \centerline{\includegraphics[width=16cm]{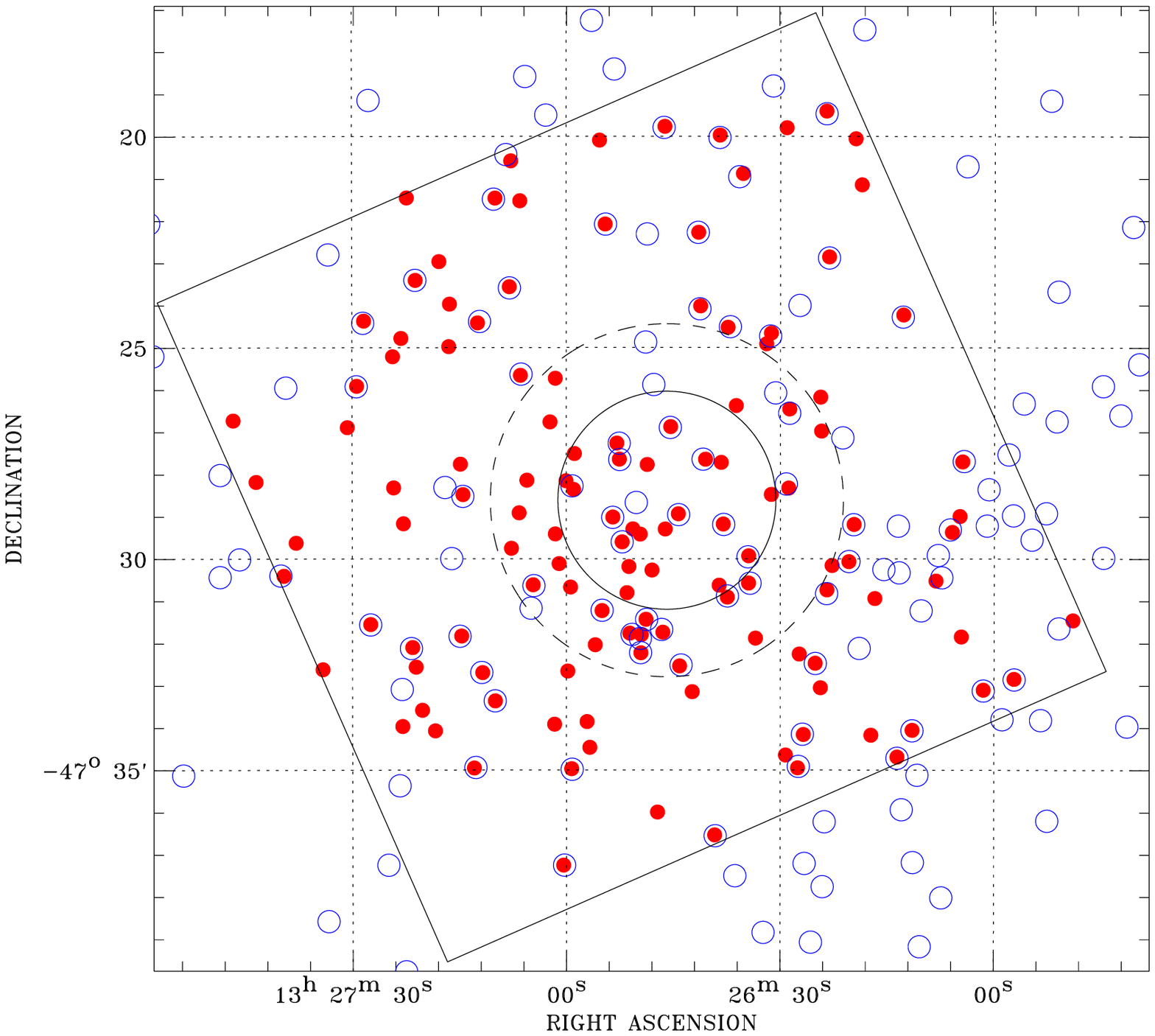}}
  \caption[]{The Chandra field of view (delimited by the square) overlaid with the \xmm\ image. The blue open circles denote \xmm\ sources and the red points indicate the Chandra sources. The sources detected both by \xmm\ and Chandra appear as red points surrounded by a blue circle. The core and half mass radii are indicated with a solid and dashed line circle respectively.}
  \label{gendre_fig4}
\end{figure*}

\begin{figure*}[!t]
  \hspace*{-0.5cm}\includegraphics[width=9cm]{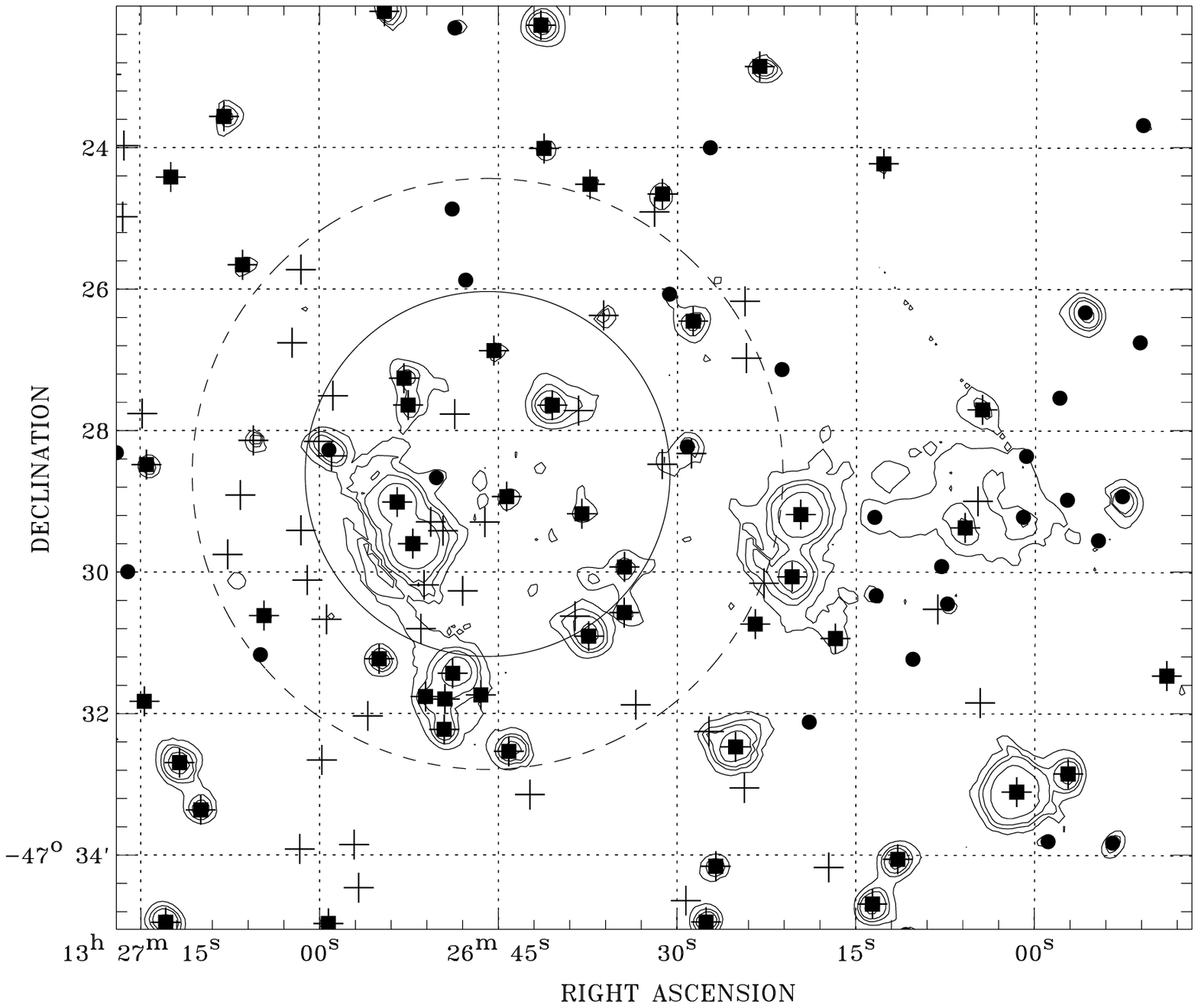}
  \hspace*{-0.5cm}\includegraphics[width=10.5cm]{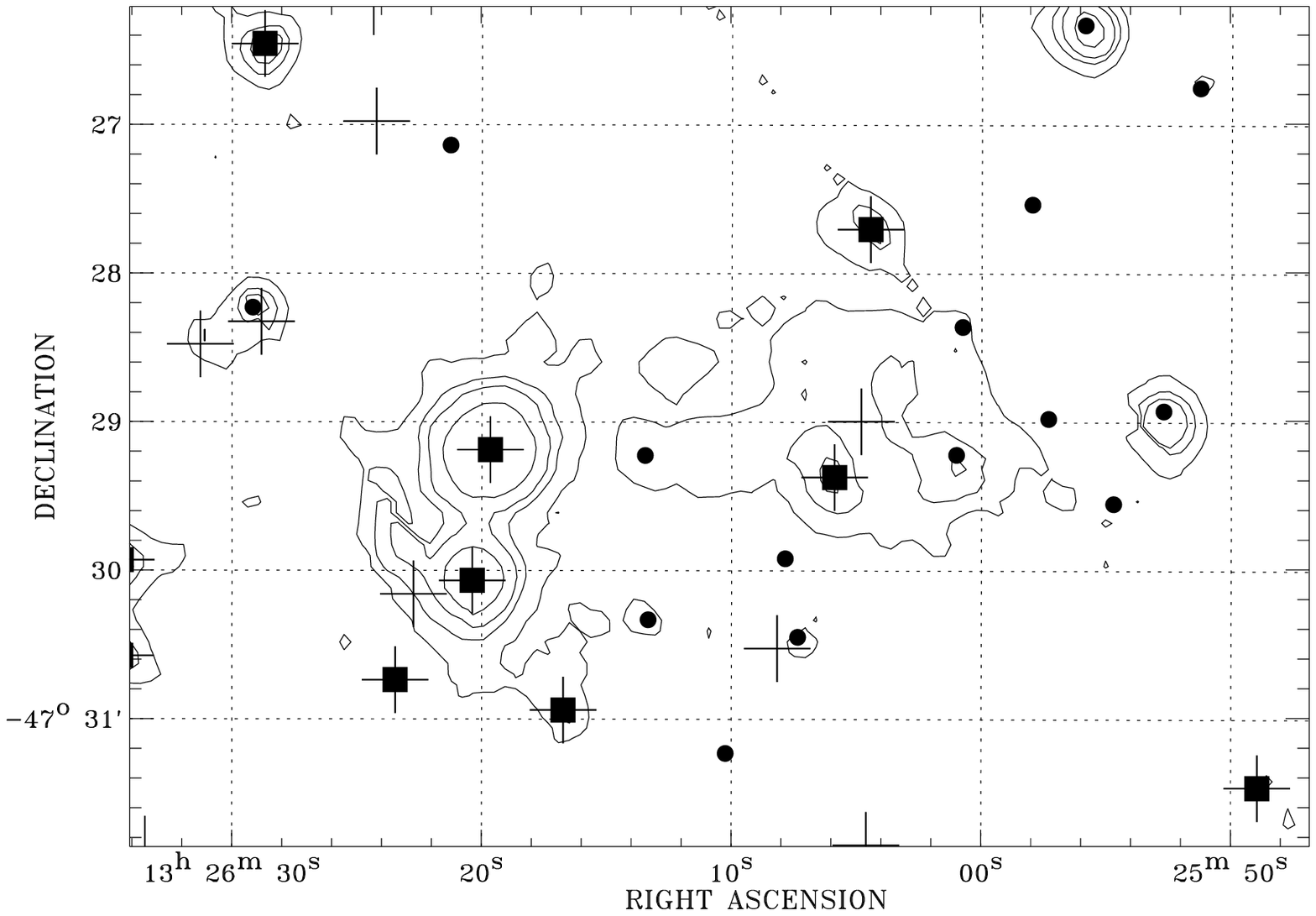}
  \caption[]{The contour images of the extended emission. The image on the left provides an overview of the cluster region. The two large circles indicate the core radius (solid line circle) and the half mass radius (dashed line circle). The Chandra sources are indicated by crosses. The right image provides a zoom on the extended emission. The \xmm\ sources are indicated by filled squares if the source is also detected by Chandra, or a filled circle otherwise. For both these images, the contour levels are 4.5, 5, 6, 7 and 8 $\sigma$ levels.}
\label{gendre_fig5}  
\label{gendre_fig6}
\end{figure*}

However, this can be improved by using the most accurate positions provided by Chandra. For this purpose, we have analyzed the publicly available Chandra observation. The Chandra observation took place on 2000 January 24-25, in imaging mode using the ACIS-I detector placed in the focus of the telescope. The observation was $\approx$ 70 kiloseconds long. These data have already been presented by \citet{coo02} and \citet{rut02}

We retrieved the data from the archives and used CIAO version 2.2.1 and CALDB version 2.12 to calibrate the event files (using the CIAO task {\it acis\_process\_events}). During the Chandra observation, the background was low and stable. We then filtered the events file for non astrophysical events using the ASCA grades 0, 2, 3, 4, 6, and for good time intervals using the provided GTI\footnote{This is the standard choice for filtering the data; see the Chandra Proposers Observatory Guide available from the Chandra web site for details.}. We also rejected events with energies below 0.2 keV. We made a basic detection scheme, using the CIAO task {\it wavdetect}. As recommended in the CIAO detect manual, we used a conservative spurious detection threshold of $10^{-5}$ as input to {\it wavdetect}. As for \xmm, sources were searched in the 0.5-5.0 keV range.

We found 129 sources within the ACIS-I field of view, to be compared to 146 in \citet{coo02} \citep[][used a sliding box algorithm and a very conservative detection threshold and reported 40 sources only]{rut02}. Even if we take into account a 1\arcsec\ systematic error in the attitude reconstruction of Chandra, the mean error on the source positions remains very small, typically $\sim1.5$\arcsec. 

The Chandra positions were then used to compute the astrometric correction for the XMM-Newton observation. We have selected three XMM-Newton sources (sources 2, 8, and 15, see Tables \ref{gendre_table1} and \ref{gendre_table2}), properly spaced within the field of view, far away from CCD gaps and bad columns, and among the brightest sources (i.e. with a small statistical error on their position). These sources are also clearly detected by Chandra. The astrometric correction was then computed with the three reference positions provided by Chandra, using the {\it starast} Interactive Data Language tools of the {\it astrolib} library. The positions listed in Tables \ref{gendre_table1} and \ref{gendre_table2} reflect this correction. 

With this correction applied, 11 and 27 sources are detected by \xmm\ within the core and half mass radii. For comparison, Chandra detected 22 and 46 sources within the same regions.

This correction further allows us to get rid of the systematic error and to cross-correlate the Chandra and XMM-Newton source positions. The positions of the Chandra sources found within the statistical error box of the XMM-Newton sources are given in Table \ref{gendre_table3}. Sixty three \xmm\ sources have a Chandra counterpart. 
We present in Fig. \ref{gendre_fig4} the Chandra field of view and sources overlaid with the \xmm\ image. 

It is beyond the scope of this paper to investigate the error box content of each \xmm\ source, and we have focussed the present paper on sources for which a previous identification has been reported.  Two EINSTEIN sources were detected by both XMM-Newton and Chandra (sources B and C), and due to its larger field of view, \xmm\ also detected the EINSTEIN sources A and D (these ones were missed by Chandra). EINSTEIN sources A and D (\xmm\ sources 3 and 7) are associated with foreground M dwarfs \citep{coo95}. The source EINSTEIN C was resolved into sources R9a and R9b in ROSAT \citep{ver00}. These two core sources which are detected as \xmm\ sources 2 and 5 are the two CV candidates \citep{car00,coo02}. The ROSAT source R20 \citep[detected by Chandra and proposed to be a BY Dra system,][]{coo02} is not detected by \xmm. The proposed quiescent neutron star binary is detected as source 4 by \xmm. Finally, the source identified by \citet[][]{ver00} as HD116789 is detected as source 28 (see Table \ref{gendre_table3}).

We have estimated the limiting count rate of the XMM-Newton observation as the count rate needed for a detection of a source placed at a mean off-axis angle of 7.5\arcmin\ which is half the radius of the EPIC-PN field of view. In the 0.5-5.0 keV band, the limiting count rate is $1.4 \times 10^{-3}$ counts s$^{-1}$. For comparison the limiting count rate of a source on-axis is $1.0 \times 10^{-3}$ counts s$^{-1}$. For Chandra, the limiting count rate estimated with the same method is $1.5 \times 10^{-4}$ counts s$^{-1}$ for an on-axis source and $1.6 \times 10^{-4}$ counts s$^{-1}$ for a source at 7.5\arcmin.

The limiting count rates have been converted into limiting fluxes using two spectral models; a blackbody of 0.6 keV and a thermal Bremsstrahlung of 3 keV. This gives $3.8\times10^{-15}$ \ergscm\ and $3.5\times10^{-15}$ \ergscm\ respectively for XMM-Newton and $1.2\times 10^{-15}$ \ergscm\ and $1.3\times10^{-15}$ \ergscm\ for Chandra. At the distance of 5.3 kpc, these fluxes translate to 0.5-5.0 keV bolometric luminosities of  $\sim 1.3\times10^{31}$ \ergs\ and $\sim 4.2\times 10^{30}$ \ergs\ for \xmm\ and Chandra respectively for the blackbody model.

As said above \xmm\ detected 146 sources, and 63 of them have a Chandra counterpart. Of the 83 remaining XMM-Newton sources, 55 were outside the field of view of Chandra. Of the remaining 28, 2 were missed by ACIS-I because of CCD gaps and noisy columns (one is in the core). This leaves a total of 26 sources detected by XMM-Newton and not detected by Chandra. We have reprocessed the Chandra data with a less conservative spurious detection threshold ($10^{-4}$) to search for fainter objects. The above number decreases from 26 to 23.  Since the Chandra observations were more sensitive than the \xmm\ one, one needs to investigate why 26 sources seen by \xmm\ were not detected by Chandra. Part of the discrepancy resides in the presence of a region of extended emission.

\section{Region of extended emission}
\label{diffus}

\citet{har82} found a region of extended emission in the EINSTEIN images. This region extended from the core to the west and the east of the cluster up to outside the half mass radius. However, ROSAT did not confirm the presence of such emission \citep{joh94}. The XMM-Newton image presented in Fig. \ref{gendre_fig5} with the Chandra sources shows that there is a region outside the half mass radius which contains 16 point sources superposed on a residual extended emission (on the right of the image). It is contained in a circle of $\approx$100$\arcsec $ radius and centered at $\alpha$ = $13^h 26^m 05^s$, $\delta$ = $-47\degr 29\arcmin 21\arcsec$. It is also present, though with less significance in the Chandra image. A zoom on this region with the detected Chandra and \xmm\ sources is shown in Fig \ref{gendre_fig6} (right). It shows that only 6 of the 16 \xmm\ sources are also detected by Chandra. It is therefore likely that the 10 additional \xmm\ sources may not be real point sources, but instead fluctuations of the extended emission which confused the detection algorithm. This and the origin of this emission will be discussed elsewhere. If we take out these 10 sources, there remains 16 \xmm\ sources which would have been detected by Chandra; had their luminosity remained constant between the two observations. As we will show later, we have found evidence for variable sources in other regions of the images, including the cluster core.

The extended emission region seen by EINSTEIN also included the core of the cluster.  The core now contains many faint X-ray sources as shown in Fig. \ref{gendre_fig5}, and we have no evidence for any residual extended emission.

\section{Search for short and long term time variability}
\label{vari}
\subsection{Variable sources within the XMM-Newton observation}

\begin{figure}[!h]
  \centerline{\includegraphics[width=9.5cm]{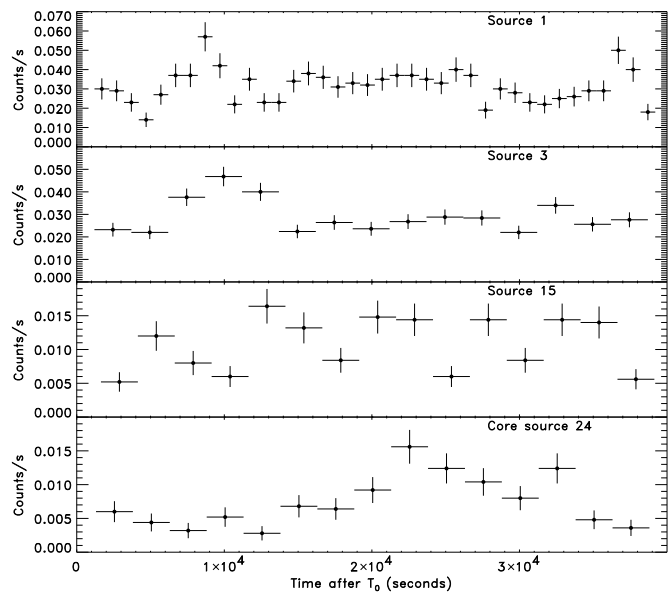}}
  \caption[]{The light curves of the four variable sources found within the field of view.  {\it Top)} The brightest object in the field (source 1). {\it 2nd)} The star USNO-A2 0375-18249604 identified by \citet{coo95} (source 3). {\it 3rd)} An object which is located within twice the half mass radius (source 15). {\it 4th)}. The only core source which showed variability within the XMM-Newton observation (source 24).}
 \label{gendre_fig7}
\end{figure}

We have extracted the light curves of the 26 strongest X-ray sources within the EPIC-PN field of view. The bin time of the light curve was chosen to ensure a sufficient number of count within each bin (typically larger than $\sim 20$). The light curves were searched for variability using the {\it lcstat} FTOOLs. Using a $\chi^2$ test, we found that 4 sources have a probability of being constant of less than 0.1\%. The light curves of these 4 objects are shown in Fig \ref{gendre_fig7}. Only one lies in the core (source 24). The three others lie outside the half mass radius.

\subsection{Variable sources between the Chandra and \xmm\ observations}

As shown above, all \xmm~sources should have been detected by Chandra, providing that their X-ray intensity had remained constant. In the core this is the case, as the only \xmm\ source not detected by Chandra fell on an ACIS-I CCD gap. Within the half mass radius, 4 \xmm\ sources were not detected by Chandra and must have therefore varied by at least a factor of $\sim 3-4$. 

We have converted the count rate of the Chandra sources to \xmm\ count rates for comparison with our count rate detection threshold. We have found that 7 Chandra sources that lie within the half mass radius should have been detected by \xmm\ and were not (this number rises to 20 if one considers the whole field of view). Finally, within the half mass radius there are 4 sources (13 in the whole field of view) detected by both Chandra and \xmm\ but with different luminosities (a factor of two or higher variations). These sources are listed in Table \ref{gendre_table4}. From this we conclude that 15 sources contained in the half mass radius have shown variability between the \xmm\ and Chandra observations spaced by $\sim 1.5$ year.

\subsection{Variable sources between the \xmm\ and ROSAT observations}

\omecen~was observed between August 1992 and January 1997 by ROSAT \citep{ver00}. The luminosity limit of the ROSAT observations was about $7\times10^{31}$ \ergs~in the 0.5-2.5 keV range \citep{ver00}. We have computed the luminosity of the \xmm\ sources in this same energy band. From this, we found that 1 source (source 13) should have been detected by ROSAT. Obviously, \xmm\ which is more sensitive than ROSAT  should have detected all ROSAT sources. This is not the case, as 1 ROSAT core source (source R20) is not present in the \xmm\ image (it is however detected by Chandra).

\section{Spectral analysis of the brightest objects}
\label{xmm_spec}

We have extracted spectra for the brightest sources; those with a total number of counts exceeding $\sim 100$ in the most sensitive EPIC-PN camera. There are 26 sources satisfying this criterion. In this paper, we limit the spectral analysis to the 16 sources lying within twice the half mass radius.

To accumulate spectra, we chose an extraction radius of $\approx$ 0.7 \arcmin, except when another source was closer than 1.5 \arcmin (two extraction radii). We extracted the background using an adjacent area of the same surface, at the same off axis angle on the same CCD. We generated ancillary response files and redistribution matrix files with the SAS tasks {\it arfgen} and {\it rmfgen} of the 5.3.3 release. 

Whenever possible, we binned the spectra to contain at least twenty net counts in each bin, in order to use $\chi^2$ statistics. Otherwise we used the Cash statistics. For the spectra with the largest number of counts, we have left the interstellar column density as a free parameter of the fit. Note however that in all but one case, the fitted \nh\ is consistent within error bars with the value expected from the optical extinction in the direction of the cluster. We used XSPEC v11.1 \citep{arn96} to fit the spectra. The limited statistics does not allow us to use spectral models more sophisticated than thermal Bremsstrahlung, blackbody, and power law.   

\subsection{Sources within the core and the half mass radii}
There are three sources in the core for which a spectrum can be extracted (sources 2, 5 and 24), two more between the core and half mass radii (sources 9 and 20) and the proposed quiescent neutron star binary which is just at the border of the half mass radius (source number 4).

\begin{figure}[!b]
  \centerline{\includegraphics[width=9.5cm]{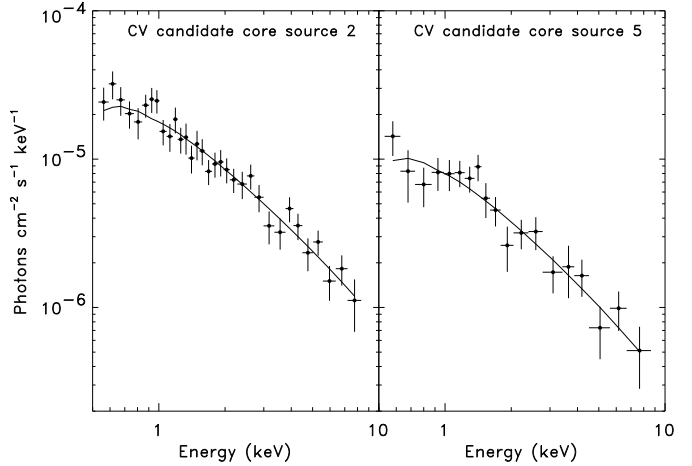}}
  \caption[]{The EPIC-PN unfolded spectra of the two CV candidates. These two sources are located within the core radius. These spectra are shown with a best thermal Bremsstrahlung  fit. The spectra measured by \xmm\ strengthen the CV classification for these two sources. These are to date the highest quality spectra of faint core globular cluster sources.}
 \label{gendre_fig8}
\end{figure}

The three core sources are the two proposed CV candidates \citep{car00} and source 24 which was found to be variable within the \xmm\ observation (see section \ref{vari} and Fig \ref{gendre_fig6}). The best fit spectral results are listed in Table \ref{gendre_table5}. For the two CV candidates, the spectra can be accurately fitted with a thermal Bremsstrahlung (or alternatively with power laws of index $\sim 1.4$). Such spectra are expected  from such systems \citep{ric96}. Thus our spectral observations reinforce the classification of these two objects as CVs. Their unfolded spectra are shown in Fig. \ref{gendre_fig8}. These are to date the highest quality spectra ever measured from  faint globular cluster X-ray sources.

For the third variable source, its spectrum is also consistent with a power law, but given the limited statistics it could be also fitted with a thermal bremsstrahlung. Despite the source faintness, we have searched for spectral variations within the observation. Two spectra were extracted, one during its steady state and another one during the flaring state. As can be seen in Table \ref{gendre_table5}, the two spectra are consistent. Best fit results for the two additional sources found between the core and half mass radii are also listed in Table \ref{gendre_table5}.

\subsection{The quiescent neutron star binary}

\begin{figure}[!b]
  \centerline{\includegraphics[width=9.5cm]{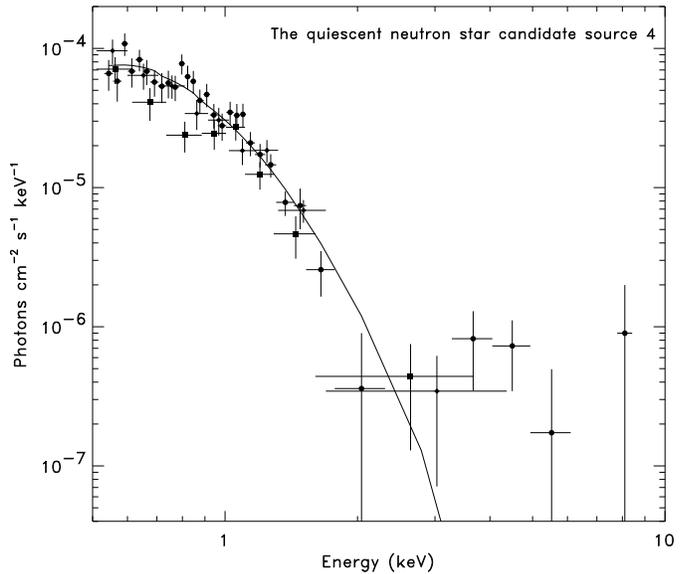}}
  \caption[]{The unfolded EPIC-MOS and EPIC-PN spectra of the quiescent neutron star binary candidate source 4. This spectrum is accurately fitted with a pure hydrogen neutron star atmosphere model. The filled squares, diamonds and circles refer to the MOS1, MOS2 and PN data respectively.}
 \label{gendre_fig9}
\end{figure}

\citet{rut02} showed that the Chandra spectrum of the quiescent neutron star binary candidate can be fitted with a pure hydrogen neutron star atmosphere model  \citep{pav92,zav96}. This source is also clearly detected in our observation, as the fourth brightest source in the field of view. Taking advantage of the better statistics of the \xmm\ spectrum, we have also fitted its spectrum with the same neutron star atmosphere model. The parameters of the latter model are the temperature, the radius, the mass of the neutron star and its distance. The temperature and the radius were derived as measured by an observer at infinity. The best fit result with the mass of the neutron star and the source distance frozen are listed in Table \ref{gendre_table6}. There is a remarkable consistency between the results reported in \citet{rut02} and ours, but thanks to the improved statistics the error bars on the fitted parameters are much smaller with the present data. We did not find any evidence for the presence of a power law tail. Assuming a power law with photon index of 2, an upper limit of 10 \% of the total flux (90\% confidence limit) can be derived for such a power law tail. The 0.1-5.0 keV bolometric luminosity measured by \xmm\ is consistent within error bars with the Chandra value. The unfolded combined EPIC-PN and MOS spectrum is presented in Fig. \ref{gendre_fig9}. This is one of the best spectra of a quiescent neutron star binary obtained so far.  Our observation thus strengthens the quiescent neutron star binary hypothesis for this object.

\subsection{Notes on remaining sources within the field of view}

\begin{figure}[!hbt]
  \centerline{\includegraphics[width=9.5cm]{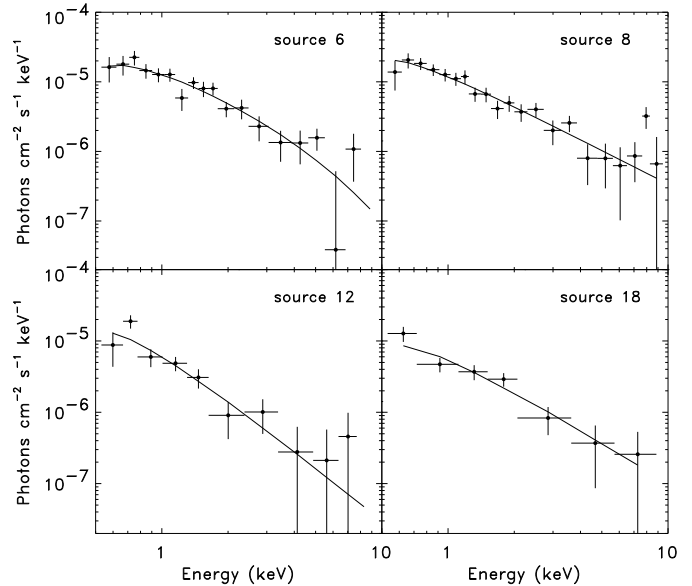}}
\caption[]{The unfolded spectra of four of the brightest objects in the EPIC-PN field of view. These sources are located outside the half mass radius, but within twice this radius. The spectrum of source 6 is shown with a thermal Bremsstrahlung fit, whereas for the others their spectra are shown with a power law fit. The four spectra are relatively hard and are consistent with those observed from the two core CVs, both in shapes and luminosities.}
 \label{gendre_fig10}
\end{figure}

In Fig. \ref{gendre_fig10}, we show the unfolded spectra of four more sources whose positions are between one and two half mass radii. All spectra are relatively hard, corresponding to colors in Fig  \ref{gendre_fig3} similar to the colors measured from the CVs. Their luminosities, just around $\sim 10^{32}$ are also consistent with the CV hypothesis. Note however that some of them may also be background sources.

\begin{figure}[!htb]

\vspace*{2cm}\centerline{\includegraphics[width=9.0cm]{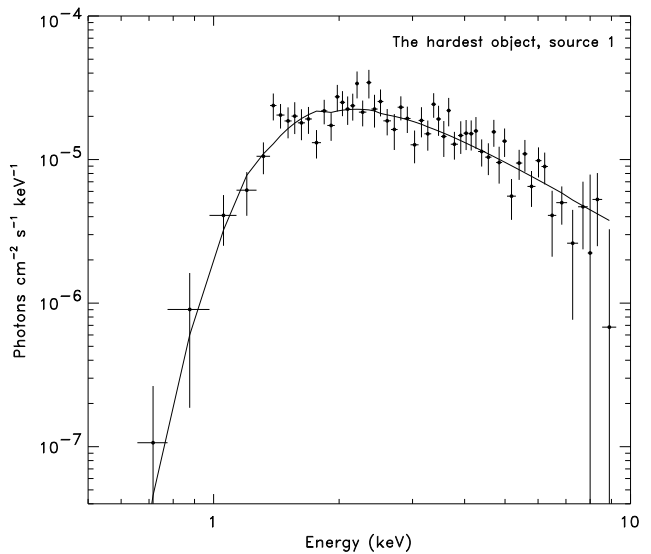}}
  \caption[]{The unfolded spectrum of the brightest object within the field of view. This source is very absorbed. Its spectrum is presented with a power law fit. It is presently unidentified, but clearly deserves some follow-up investigations.}
 \label{gendre_fig11}
\end{figure}

In addition, there are two other sources in the field of view which deserve some attention. The first one is the  brightest object (it was detected by both Chandra and ROSAT, see Table \ref{gendre_table3}). Its unfolded spectrum together with its best power law fit is presented in Fig. \ref{gendre_fig11}. The high \nh\ derived from the fit and its relatively large angular distance from the cluster center (about 8.5\arcmin) calls into question its membership to \omecen. However, its unusual properties (time variability and hard spectrum) make it an interesting target for follow-up investigations. There are no counterparts listed in the USNO A2.0 catalog within 2" of the Chandra position.

The second object is the third brightest object in the field (source 3). It has an M dwarf counterpart \citep[USNO-A2 0375-18249604, ][]{coo95}. Its spectrum is well fitted by a 2 temperature Raymond-Smith model (2T) expected from such a system \citep{sin96}. The best fit parameters of all these objects are also given in Table \ref{gendre_table5}.

\section{Discussion}
\label{discu}

We have presented the first results of our deep \xmm\ observation of the globular cluster \omecen, emphasizing the general properties of the population of faint X-ray sources present in the cluster. We have detected 11 and 27 faint X-ray sources within the core and half-mass radii respectively. We have estimated that $4\pm1$ and $9\pm2$ of these objects could be unrelated background sources. Comparing the Chandra ACIS-I and \xmm\ EPIC observations, we have found that 63 \xmm\ sources have a Chandra counterpart. Fifteen sources that lie within the half mass radius have shown variability between the two observations. Intensity variations for several sources were also found within the \xmm\ observation and between the \xmm\ and previous ROSAT observations. We have also presented the first X-ray spectra of the brightest and peculiar objects in the field, in particular the proposed quiescent neutron star low mass X-ray binary for which the EPIC spectrum strengthens its classification. We have shown that the spectra of the two brightest core sources strongly support the proposal that they are CVs. We have found objects with similar spectra in the cluster. In the following, we briefly discuss what appears to be the main implications of our observation. 

First of all, what is striking from our data is the excess of sources located in the vicinity of the cluster (just outside the half mass radius). This has already been noted by \citet{coo02} and \citet{ver02}.  For instance between 1 and 2 half-mass radii, there is an excess of $\sim 25$ sources, over the expected number of background sources. Obviously, some of them may be foreground stars. However, some may also belong to the cluster, and may have already been found as active binaries (RS CVn). Two of the  OGLEGC sources located at relatively large off-axis (more than 2.5 core radii) were detected by Chandra \citep[OGLEGC15 and OGLEGC22,][]{coo02}. Another one (OGLEGC30) has a position coincident with our \xmm\ source 29. All are proposed to be RS CVn stars \citep{kal96, kal02}. Their X-ray luminosity is not atypical of such systems $\sim 1-3 \times 10^{31}$ \ergs, though on the bright end of their luminosity distribution \citep{dem97}.

Due to mass segregation effects, binaries which are more massive than lonely stars are expected to lie close to cluster center \citep[e.g. ][]{mey97}. In \omecen\, the mass segregation is very low, and that might explain the large population of binaries outside the core \citep[e.g.][]{ver00}. Some of these binaries might also have been ejected outside the cluster through three-body interactions (an encounter of a binary with a single star). Another possibility could be that the potential well of the cluster was recently disrupted by the accretion of a discrete component of another stellar system. The recent discovery of a metal-rich stellar population in the cluster \citep{pan00} with a coherent bulk motion with respect to the other stars \citep{fer02} was explained by the accretion of an independent stellar system by \omecen. The presence of faint X-ray sources in the vicinity of \omecen\ might thus be a consequence of an unusual dynamical evolution of the cluster.

The second most striking feature of our data is the lack of soft X-ray sources and the large number of sources showing long term variability. In Fig. \ref{gendre_fig2} and \ref{gendre_fig3}, there is an obvious clustering of sources around the two previously identified CVs (which are spectrally hard). There are only two sources in the soft area: the star identified by \citet{coo95} and the proposed quiescent neutron star binary (note that these two objects are also among the brightest, see Fig. \ref{gendre_fig3}). The sources clustering below the two CV candidates (sources 2 and 5, represented by two filled squares) in the hardness intensity plot (see Fig. \ref{gendre_fig3}) have luminosities in the range  $2 \times 10^{31}$ to $6 \times 10^{32}$ ergs s$^{-1}$. The spectra of four of these objects are shown in Fig. \ref{gendre_fig9}. Similar spectra and luminosities are observed from disk and globular cluster CVs \citep[e.g. ][]{poo02a}. CVs are well known to be variable, and might thus account for some of the variable sources present in the cluster (note that one core CV, source 5, showed variability by a factor of $\sim 1.7$ between the \xmm\ and Chandra observations).  

However, we note that one of the proposed RS CVns \citep[OGLEGC30,][ the counterpart of the \xmm\ source 29]{kal96} has colors similar to the two CVs (the statistic was unfortunately too poor to fit its spectrum). This source  was not detected by Chandra and must have varied by at least a factor of $\sim 3-4$ between the two observations. The two OGLEGC sources (RS CVn candidates) detected by Chandra were not detected by \xmm: the Chandra luminosity of OGLEGC 15 was below the \xmm\ sensitivity threshold, on the other hand OGLEGC 22 should have been detected. Another example of variability is given by the ROSAT source R20, associated with a BY Dra \citep{coo02}. This source was detected by Chandra but not by \xmm. These objects are variable, and will be preferentially detected during flaring outbursts, due to their low quiescent X-ray luminosities. The large number of variable sources in \omecen\ (15 within the half mass radius between the Chandra and XMM-Newton observations) is also suggestive of a large population of RS CVns (including BY Dra) in the cluster. During snapshot X-ray observations, only a fraction could be seen. These sources could account for the population of the lower luminosity ($\sim 10^{31}$\ergs) sources found in \omecen.

Some of the faint and persistent sources of the hardness-intensity diagram ($\sim 10$, see Fig. \ref{gendre_fig3}) have colors consistent with power law like spectra, with indices of the order of $\sim 2$. Such power laws could result from magnetospheric emission of millisecond pulsars \citep{bec98, web02b}.  However, these sources are unlikely to be millisecond pulsars, because no such radio pulsars are presently known in the cluster \citep{fre02}. Furthermore, \citet{gri02} have recently shown that the emission of the millisecond pulsars detected in 47 Tuc is dominated by the thermal emission from the polar caps of the neutron star. Such emission is much softer than the magnetospheric emission. These sources with soft thermal X-ray emission should lie on figure \ref{gendre_fig3} between vertical lines passing through the two CV candidates and the quiescent neutron star binary. There are no sources in that region. Our observation would thus support the idea that \omecen\ lacks millisecond pulsars. Obviously the difficulties in retaining neutron stars in a globular cluster and the low collision frequency of \omecen\ could provide an explanation \citep[see e.g.][]{pfa02,ver02}.

In the disk, quiescent neutron star binaries have luminosities in the range $10^{32}-10^{33}$ erg s$^{-1}$ \citep{nar02}. Furthermore, they all have extremely soft X-ray spectra \citep{rut00}. In our observation, there is one single object with these characteristics. If globular cluster quiescent neutron star binaries behave similarly to those in the disk, then our observation should provide a complete census of the content of such objects in \omecen\ \citep[a similar conclusion was derived by][ from the Chandra observations]{rut02}. Some of these objects have also been found in other clusters \citep{edm02,gri01a,poo02b}. In globular clusters, these systems are certainly formed from a close encounter between a neutron star with a single star or with a binary \citep[see][ for a recent review]{ver02b}. The  presence of one such system in \omecen\ is consistent with its lower collision frequency compared to other clusters \citep{ver02b}. The presence of this object far away from the core remains somewhat puzzling in that regard. It might have been ejected from the core during a three-body interaction \citep{ver02b}. The presence of a single candidate quiescent neutron star binary together with the apparent lack of millisecond pulsars in the cluster makes \omecen\ clearly different from clusters like 47 Tuc in which a large population of binaries with neutron star primaries are being discovered \citep[e.g.][]{gri01a,cam00}. This difference, if confirmed, should help us in understanding how neutron stars form and evolve in globular clusters. 

Finally, some of the faint sources could be quiescent black hole binaries \citep[one black hole may have just been discovered in the globular cluster M15, and some are being found in globular clusters in other galaxies,][]{ger02,ver02b}. Most quiescent black hole binaries within the galactic disk have been observed with luminosities of $\sim 10^{31}$ erg s$^{-1}$ \citep{kon02,ham02}. Their spectra are well fitted with power laws, with spectral indices of 1.5-2 \citep{kon02}. Such spectra correspond to hardness ratios in the range 0.4-0.9 in Fig. \ref{gendre_fig3}. About 10 objects have colors and luminosities consistent with a quiescent black hole binary nature. This hypothesis is however poorly constrained from our X-ray observations alone, as many faint sources could be background active galactic nuclei.

\section{Conclusions}
\label{conclu}
The main results of our XMM-Newton observation are that in Omega Cen the binaries do not seem be confined to the core or even within the half mass radius, as previously thought, and that the majority of the faint X-ray sources may be CVs, RS CVn or BY Dra binaries. Obviously, the small number of identifications currently available prevents us from reaching definite conclusions. The spectral and timing information provided by XMM-Newton, together with the accurate positions provided by Chandra for a wide sample of faint X-ray sources should encourage follow-up investigations at other wavelengths (optical, UV, radio). In particular, these observations should tell us whether those sources found well outside the core of \omecen\ really belong to the cluster. More identifications are also required before reliable comparisons between \omecen\ and other clusters with different structural parameters can be drawn.  Such comparisons are critical to a better understanding of the dynamical evolution of globular clusters in general.

\begin{acknowledgements}
The authors would like to thank A. Cool and F. Verbunt for helpful discussions and the referee, H. Johnston for valuable comments. We would also like to thank M. Watson and P. Marty for their help during the analysis of the extended emission region, V. E. Zavlin for providing us with the pure H neutron star atmosphere model, and G. Hasinger for discussions about the interpretation of the Lockman Hole results. Finally, we are grateful to G. Lamer and J. Tedds for help with the astrometric correction.
\end{acknowledgements}

\def\baselinestretch{1.0}
\begin{table*}[!t]
\caption{XMM-Newton sources detected within the half mass radius. A flag indicates whether the source belongs to the core. Most sources were detected by the most sensitive EPIC-PN camera (a flag indicates whether it is detected also by MOS1 and MOS2). The ones which were missed by EPIC-PN are labeled as MOS in that table. The source ID increases with decreasing maximum likelihood. The sources are sorted in declination. The statistical error on the position is given at the 90\% confidence level, as estimated from the {\it srcmatch} task, and is given in arcseconds. For PN sources, the rates are from the PN data. For MOS sources, the rates are an average between the MOS1 and MOS2 cameras. All rates are corrected for the vignetting of the mirrors and are given in units of $10^{-2}$ count s$^{-1}$. In EPIC-PN, for indications a count rate of $10^{-2}$ counts s$^{-1}$ corresponds to an unabsorbed flux of $\sim 2.7\times10^{-14}$ \ergscm\ for a 0.6 keV blackbody model absorbed through the cluster \nh\ ($8.4\times 10^{20}$ cm$^{-2}$). This flux translates to a luminosity of $\sim 10^{32}$ \ergs\ at the distance of \omecen\ (5.3 kpc). 
\label{gendre_table1}}
\begin{center}
\begin{tabular}{ccccccccc}
Instrument & Source &  RA   & Dec   & Position & Rate  &  MOS1 & MOS2 & Core\\
 & ID  &   h  m  s  & $\degr$ $\arcmin$ $\arcsec$ & Error (\arcsec)  &    & detection & detection & source\\

\noalign{\smallskip\hrule\smallskip}

PN &  122 &              13: 26: 48.9  & -47: 24: 52.2 &   7.32 & 0.13 $\pm$ 0.03 &  No &  No &  No \\
PN &   80 &              13: 26: 30.7  & -47: 26: 04.4 &   4.76 & 0.14 $\pm$ 0.03 & Yes &  No &  No \\
PN &   49 &              13: 26: 28.7  & -47: 26: 33.3 &   5.28 & 0.15 $\pm$ 0.03 &  No & Yes &  No \\
PN &  103 &              13: 26: 45.4  & -47: 26: 53.6 &   4.92 & 0.15 $\pm$ 0.03 &  No &  No & Yes \\
PN &   44 &              13: 26: 52.6  & -47: 27: 15.9 &   3.89 & 0.25 $\pm$ 0.04 & Yes & Yes & Yes \\
PN &   24 &              13: 26: 40.8  & -47: 27: 38.6 &   2.35 & 0.54 $\pm$ 0.05 & Yes & Yes & Yes \\
PN &   61 &              13: 26: 52.5  & -47: 27: 39.1 &   4.08 & 0.18 $\pm$ 0.03 & Yes & Yes & Yes \\
PN &   42 &              13: 26: 59.2  & -47: 28: 16.5 &   4.41 & 0.35 $\pm$ 0.04 & Yes & Yes & Yes \\
PN &  111 &              13: 26: 50.2  & -47: 28: 40.1 &   4.47 & 0.14 $\pm$ 0.03 &  No &  No & Yes \\
PN &   92 &              13: 26: 44.3  & -47: 28: 57.4 &   4.89 & 0.17 $\pm$ 0.03 &  No &  No & Yes \\
PN &    5 &              13: 26: 53.5  & -47: 29: 01.3 &   1.17 & 1.74 $\pm$ 0.08 & Yes & Yes & Yes \\
PN &  101 &              13: 26: 37.9  & -47: 29: 11.3 &   5.06 & 0.16 $\pm$ 0.03 &  No &  No & Yes \\
PN &    2 &              13: 26: 52.2  & -47: 29: 36.1 &   0.74 & 3.56 $\pm$ 0.11 & Yes & Yes & Yes \\
PN &   72 &              13: 26: 34.5  & -47: 29: 57.3 &   3.84 & 0.25 $\pm$ 0.04 &  No &  No & Yes \\
PN &   60 &              13: 26: 34.2  & -47: 30: 34.7 &   4.08 & 0.25 $\pm$ 0.04 & Yes &  No &  No \\
PN &  127 &              13: 27: 04.5  & -47: 30: 38.2 &   5.43 & 0.12 $\pm$ 0.03 &  No &  No &  No \\
PN &   20 &              13: 26: 37.4  & -47: 30: 52.9 &   2.32 & 0.53 $\pm$ 0.05 & Yes & Yes &  No \\
PN &   88 &              13: 27: 04.9  & -47: 31: 10.2 &   6.88 & 0.13 $\pm$ 0.03 & Yes &  No &  No \\
PN &   38 &              13: 26: 55.0  & -47: 31: 13.2 &   3.73 & 0.29 $\pm$ 0.04 & Yes & Yes &  No \\
PN &    9 &              13: 26: 48.7  & -47: 31: 26.2 &   1.72 & 1.46 $\pm$ 0.10 & Yes & Yes &  No \\
PN &  108 &              13: 26: 46.6  & -47: 31: 40.5 &   5.79 & 0.16 $\pm$ 0.03 &  No &  No &  No \\
PN &   36 &              13: 26: 50.8  & -47: 31: 47.5 &   3.91 & 0.33 $\pm$ 0.04 & Yes & Yes &  No \\
PN &   35 &              13: 26: 49.6  & -47: 31: 53.2 &   7.26 & 0.17 $\pm$ 0.03 & Yes & Yes &  No \\
PN &   57 &              13: 26: 49.6  & -47: 32: 13.4 &   2.95 & 0.35 $\pm$ 0.05 &  No &  No &  No \\
PN &   16 &              13: 26: 43.9  & -47: 32: 31.0 &   2.71 & 0.68 $\pm$ 0.07 & Yes & Yes &  No \\
MOS&  96 &              13: 26: 47.7  & -47: 25: 52.6 &   4.71 & 0.05 $\pm$ 0.02 & Yes & Yes &  No \\
MOS&  50 &              13: 26: 29.2  & -47: 28: 13.8 &   2.99 & 0.13 $\pm$ 0.02 & Yes & Yes &  No \\
\noalign{\smallskip\hrule\smallskip}
\end{tabular}
\end{center}
\end{table*}

\begin{table*}[!th]
\caption{XMM-Newton sources detected outside the half mass radius. As in Table \ref{gendre_table1} most sources were detected by the most sensitive EPIC-PN camera (a flag indicates whether it is detected also by MOS1 and MOS2). The ones which were missed by EPIC-PN are labeled as MOS in that table. The source ID increases with decreasing maximum likelihood. The sources are sorted in declination. The statistical error on the position is given at the 90\% confidence level, as estimated from the {\it srcmatch} task and is given in arcseconds. For PN sources, the rates are from the PN data. For MOS sources, the rates are an average between the MOS1 and MOS2 cameras. All rates are corrected for the vignetting of the mirrors and are given in units of $10^{-2}$ count s$^{-1}$. In EPIC-PN, for indications a count rate of $10^{-2}$ counts s$^{-1}$ corresponds to an unabsorbed flux of $\sim 2.7\times10^{-14}$ \ergscm\ for a 0.6 keV blackbody model absorbed through the cluster \nh\ ($8.4\times 10^{20}$ cm$^{-2}$). This flux translates to a luminosity of $\sim 10^{32}$ \ergs\ at the distance of \omecen\ (5.3 kpc). \label{gendre_table2}}
\begin{center}

\begin{tabular}{cccccccc}

Instrument & Source &  RA   & Dec   & Position & Rate  &  MOS1 & MOS2 \\
 & ID  & h  m  s  & $\degr$ $\arcmin$ $\arcsec$  & Error (\arcsec)   &   & detection & detection \\

\noalign{\smallskip\hrule\smallskip}
PN&  119 &              13: 26: 37.8  & -47: 16: 00.8 &   7.59 & 0.31 $\pm$ 0.06 &  No &  No\\
PN  &   54 &              13: 26: 56.4  & -47: 17: 15.4 &   3.77 & 0.53 $\pm$ 0.07 &  No & Yes\\
PN  &   65 &              13: 26: 18.3  & -47: 17: 28.6 &   5.31 & 0.43 $\pm$ 0.07 & Yes & Yes\\
PN  &   29 &              13: 26: 53.3  & -47: 18: 24.4 &   3.60 & 0.63 $\pm$ 0.07 & Yes & Yes\\
PN  &   97 &              13: 27: 05.8  & -47: 18: 35.1 &   5.02 & 0.30 $\pm$ 0.05 &  No &  No\\
PN  &  132 &              13: 26: 31.0  & -47: 18: 48.5 &  26.59 & 0.21 $\pm$ 0.05 &  No &  No\\
PN  &    3 &              13: 27: 27.7  & -47: 19: 08.8 &   0.83 & 7.39 $\pm$ 0.24 & Yes & Yes\\
PN  &   17 &              13: 26: 23.5  & -47: 19: 27.9 &   3.00 & 0.89 $\pm$ 0.08 & Yes & Yes\\
PN  &   56 &              13: 27: 02.8  & -47: 19: 29.9 &   5.59 & 0.38 $\pm$ 0.06 & Yes & Yes\\
PN  &   28 &              13: 26: 46.3  & -47: 19: 47.4 &   2.64 & 0.80 $\pm$ 0.07 & Yes & Yes\\
PN  &   31 &              13: 26: 38.5  & -47: 20: 01.7 &   2.43 & 0.66 $\pm$ 0.07 & Yes & Yes\\
PN  &   95 &              13: 27: 08.4  & -47: 20: 25.9 &   7.41 & 0.16 $\pm$ 0.04 & Yes &  No\\
PN  &   43 &              13: 26: 03.8  & -47: 20: 42.9 &   3.86 & 0.54 $\pm$ 0.07 & Yes & Yes\\
PN  &   73 &              13: 26: 35.7  & -47: 20: 57.4 &   9.14 & 0.26 $\pm$ 0.05 &  No & Yes\\
 PN &   68 &              13: 27: 10.2  & -47: 21: 29.2 &   4.06 & 0.31 $\pm$ 0.05 & Yes &  No\\
PN  &  145 &              13: 28: 08.7  & -47: 21: 36.8 &  48.16 & 0.54 $\pm$ 0.09 &  No &  No\\
PN  &  133 &              13: 27: 58.4  & -47: 22: 03.6 &  24.29 & 0.30 $\pm$ 0.06 &  No &  No\\
PN  &   32 &              13: 26: 54.5  & -47: 22: 04.7 &   2.73 & 0.46 $\pm$ 0.06 & Yes & Yes\\
PN  &   15 &              13: 26: 41.5  & -47: 22: 16.5 &   2.34 & 0.66 $\pm$ 0.06 & Yes & Yes\\
PN  &   77 &              13: 26: 48.6  & -47: 22: 18.7 &   4.95 & 0.15 $\pm$ 0.03 &  No & Yes\\
PN  &  129 &              13: 27: 33.3  & -47: 22: 48.1 &   6.29 & 0.18 $\pm$ 0.04 &  No &  No\\
PN  &   39 &              13: 26: 23.2  & -47: 22: 52.7 &   2.70 & 0.52 $\pm$ 0.06 & Yes &  No\\
PN  &   70 &              13: 28: 01.9  & -47: 23: 16.5 &   4.68 & 0.68 $\pm$ 0.09 &  No &  No\\
PN  &   18 &              13: 27: 21.2  & -47: 23: 24.4 &   1.92 & 0.92 $\pm$ 0.08 & Yes & Yes\\
PN  &   41 &              13: 27: 07.9  & -47: 23: 35.3 &   3.28 & 0.42 $\pm$ 0.05 & Yes & Yes\\
PN  &   99 &              13: 25: 51.0  & -47: 23: 40.6 &   5.60 & 0.33 $\pm$ 0.06 &  No &  No\\
PN  &  118 &              13: 26: 27.3  & -47: 24: 00.2 &  17.34 & 0.20 $\pm$ 0.04 &  No &  No\\
PN  &   67 &              13: 26: 41.3  & -47: 24: 05.0 &   6.67 & 0.17 $\pm$ 0.03 & Yes & Yes\\
PN  &   79 &              13: 26: 12.8  & -47: 24: 16.6 &   5.92 & 0.19 $\pm$ 0.04 &  No & Yes\\
PN  &  110 &              13: 27: 12.1  & -47: 24: 22.9 &   5.66 & 0.17 $\pm$ 0.04 &  No &  No\\
PN  &   94 &              13: 27: 28.5  & -47: 24: 25.2 &   4.86 & 0.23 $\pm$ 0.04 &  No &  No\\
PN  &  126 &              13: 26: 37.0  & -47: 24: 30.5 &   6.26 & 0.12 $\pm$ 0.03 &  No &  No\\
PN  &   59 &              13: 26: 31.4  & -47: 24: 43.6 &   5.14 & 0.24 $\pm$ 0.04 & Yes & Yes\\
PN  &   86 &              13: 27: 57.9  & -47: 25: 12.0 &   6.47 & 0.29 $\pm$ 0.06 &  No & Yes\\
PN  &   75 &              13: 25: 39.7  & -47: 25: 23.4 &   6.02 & 0.35 $\pm$ 0.06 &  No & Yes\\
PN  &   82 &              13: 27: 06.3  & -47: 25: 37.8 &   7.52 & 0.17 $\pm$ 0.03 &  No & Yes\\
PN  &  106 &              13: 25: 44.8  & -47: 25: 54.7 &   7.58 & 0.31 $\pm$ 0.06 &  No &  No\\
PN  &    6 &              13: 27: 29.4  & -47: 25: 55.4 &   1.18 & 2.35 $\pm$ 0.11 & Yes & Yes\\
PN  &  146 &              13: 27: 39.3  & -47: 25: 57.1 &   7.76 & 0.15 $\pm$ 0.04 &  No &  No\\
PN  &  144 &              13: 27: 59.4  & -47: 26: 09.5 &   7.58 & 0.22 $\pm$ 0.05 &  No &  No\\
PN  &   26 &              13: 25: 55.9  & -47: 26: 19.6 &   2.53 & 0.85 $\pm$ 0.08 & Yes & Yes\\
PN  &  131 &              13: 25: 42.3  & -47: 26: 36.1 &   9.30 & 0.25 $\pm$ 0.05 &  No &  No\\
PN  &  105 &              13: 25: 51.3  & -47: 26: 44.9 &   5.92 & 0.26 $\pm$ 0.05 &  No &  No\\
PN  &  115 &              13: 26: 21.3  & -47: 27: 08.2 &   4.93 & 0.14 $\pm$ 0.03 &  No &  No\\
PN  &  138 &              13: 25: 58.0  & -47: 27: 32.0 &   6.95 & 0.16 $\pm$ 0.04 &  No &  No\\
PN  &  135 &              13: 25: 32.6  & -47: 27: 37.8 &   8.22 & 0.25 $\pm$ 0.06 &  No &  No\\
 PN &   76 &              13: 26: 04.2  & -47: 27: 41.6 &   7.24 & 0.15 $\pm$ 0.04 &  No & Yes\\
PN  &  140 &              13: 27: 48.5  & -47: 28: 01.0 &  47.60 & 0.20 $\pm$ 0.04 &  No &  No\\
PN  &  102 &              13: 27: 17.0  & -47: 28: 18.7 &   5.41 & 0.12 $\pm$ 0.03 & Yes &  No\\
PN  &  136 &              13: 26: 00.8  & -47: 28: 21.5 &  10.83 & 0.17 $\pm$ 0.04 &  No &  No\\

\end{tabular}

\end{center}

\end{table*}

\setcounter{table}{1}

\begin{table*}[!th]

\caption{Continued}
\begin{center}

\begin{tabular}{cccccccc}

Instrument & Source &  RA   & Dec   & Position & Rate  &  MOS1 & MOS2 \\
 & ID  &  h  m  s  & $\degr$ $\arcmin$ $\arcsec$  & Error (\arcsec)  &  & detection & detection \\

PN  &   47 &              13: 27: 14.5  & -47: 28: 31.3 &   4.40 & 0.28 $\pm$ 0.04 & Yes & Yes\\
PN  &   33 &              13: 25: 52.7  & -47: 28: 55.4 &   3.11 & 0.71 $\pm$ 0.08 & Yes & Yes\\
PN  &   98 &              13: 25: 57.3  & -47: 28: 58.6 &   6.54 & 0.27 $\pm$ 0.05 &  No &  No\\
PN  &    4 &              13: 26: 19.7  & -47: 29: 11.5 &   0.78 & 3.42 $\pm$ 0.12 & Yes & Yes\\
PN  &   91 &              13: 26: 13.5  & -47: 29: 13.5 &   6.09 & 0.21 $\pm$ 0.04 &  No &  No\\
PN  &   62 &              13: 26: 01.0  & -47: 29: 13.2 &   5.70 & 0.31 $\pm$ 0.05 &  No & Yes\\
PN  &   51 &              13: 26: 06.2  & -47: 29: 18.7 &   5.21 & 0.53 $\pm$ 0.06 &  No &  No\\
PN  &   64 &              13: 28: 14.1  & -47: 29: 24.0 &   5.99 & 0.92 $\pm$ 0.10 &  No &  No\\
PN  &  123 &              13: 25: 54.7  & -47: 29: 32.9 &  11.31 & 0.20 $\pm$ 0.04 &  No &  No\\
PN  &   78 &              13: 26: 07.9  & -47: 29: 55.1 &   4.43 & 0.27 $\pm$ 0.04 &  No &  No\\
PN  &   27 &              13: 25: 44.7  & -47: 29: 58.5 &   3.42 & 0.67 $\pm$ 0.08 & Yes & Yes\\
PN  &  107 &              13: 27: 16.1  & -47: 29: 59.7 &   5.76 & 0.17 $\pm$ 0.04 &  No &  No\\
PN  &  100 &              13: 27: 45.8  & -47: 30: 00.8 &   7.85 & 0.17 $\pm$ 0.04 & Yes &  No\\
PN  &  137 &              13: 26: 15.5  & -47: 30: 15.1 &  47.94 & 0.13 $\pm$ 0.03 &  No &  No\\
PN  &  104 &              13: 26: 13.3  & -47: 30: 19.8 &   5.20 & 0.18 $\pm$ 0.04 &  No &  No\\
PN  &   84 &              13: 27: 40.0  & -47: 30: 24.0 &   5.91 & 0.22 $\pm$ 0.04 & Yes &  No\\
PN  &  109 &              13: 27: 48.5  & -47: 30: 26.0 &   4.44 & 0.23 $\pm$ 0.05 &  No &  No\\
PN  &   87 &              13: 26: 07.4  & -47: 30: 26.8 &   5.88 & 0.25 $\pm$ 0.04 &  No &  No\\
PN  &  117 &              13: 26: 23.5  & -47: 30: 49.4 &   6.53 & 0.15 $\pm$ 0.03 &  No &  No\\
PN  &  124 &              13: 26: 10.2  & -47: 31: 13.8 &  27.02 & 0.16 $\pm$ 0.04 &  No &  No\\
PN  &   12 &              13: 27: 27.4  & -47: 31: 33.7 &   2.14 & 0.93 $\pm$ 0.08 & Yes & Yes\\
PN  &  121 &              13: 25: 50.9  & -47: 31: 38.9 &  25.26 & 0.23 $\pm$ 0.05 &  No &  No\\
PN  &  114 &              13: 27: 14.9  & -47: 31: 50.0 &   5.49 & 0.15 $\pm$ 0.03 &  No &  No\\
PN  &   74 &              13: 27: 21.7  & -47: 32: 07.3 &   7.32 & 0.15 $\pm$ 0.04 &  No & Yes\\
PN  &  139 &              13: 26: 18.9  & -47: 32: 07.2 &  15.62 & 0.28 $\pm$ 0.04 &  No &  No\\
PN  &    8 &              13: 26: 25.1  & -47: 32: 28.7 &   1.12 & 2.10 $\pm$ 0.10 & Yes & Yes\\
PN  &   11 &              13: 27: 11.9  & -47: 32: 41.4 &   1.86 & 0.82 $\pm$ 0.07 & Yes & Yes\\
PN  &   21 &              13: 25: 57.2  & -47: 32: 51.6 &   2.34 & 0.86 $\pm$ 0.08 & Yes & Yes\\
PN  &  116 &              13: 27: 23.0  & -47: 33: 05.4 &   7.07 & 0.18 $\pm$ 0.04 &  No &  No\\
PN  &    1 &              13: 26: 01.5  & -47: 33: 07.4 &   0.69 & 6.94 $\pm$ 0.20 & Yes & Yes\\
PN  &   34 &              13: 27: 10.0  & -47: 33: 21.8 &   3.34 & 0.36 $\pm$ 0.05 & Yes & Yes\\
PN  &   14 &              13: 28: 09.1  & -47: 33: 27.0 &   2.42 & 2.39 $\pm$ 0.16 & Yes &  No\\
PN  &  130 &              13: 25: 58.9  & -47: 33: 48.2 &  46.71 & 0.22 $\pm$ 0.05 &  No &  No\\
PN  &   37 &              13: 25: 53.5  & -47: 33: 49.3 &   3.58 & 0.62 $\pm$ 0.07 & Yes & Yes\\
PN  &   90 &              13: 25: 41.3  & -47: 33: 58.0 &   8.28 & 0.28 $\pm$ 0.06 & Yes &  No\\
PN  &   25 &              13: 26: 11.5  & -47: 34: 04.2 &   2.43 & 0.78 $\pm$ 0.07 & Yes & Yes\\
PN  &   40 &              13: 26: 26.9  & -47: 34: 09.1 &   3.88 & 0.33 $\pm$ 0.05 & Yes & Yes\\
PN  &   22 &              13: 26: 13.6  & -47: 34: 43.2 &   2.44 & 0.88 $\pm$ 0.08 & Yes & Yes\\
PN  &   30 &              13: 26: 27.4  & -47: 34: 54.9 &   2.94 & 0.50 $\pm$ 0.06 & Yes & Yes\\
PN  &   23 &              13: 27: 12.7  & -47: 34: 56.4 &   2.38 & 0.77 $\pm$ 0.07 & Yes & Yes\\
PN  &   52 &              13: 26: 59.2  & -47: 34: 59.0 &   3.58 & 0.31 $\pm$ 0.05 & Yes & Yes\\
PN  &   93 &              13: 26: 10.8  & -47: 35: 07.3 &   4.89 & 0.27 $\pm$ 0.05 &  No &  No\\
PN  &   85 &              13: 27: 53.7  & -47: 35: 07.6 &   6.17 & 0.28 $\pm$ 0.06 & Yes &  No\\
PN  &  141 &              13: 27: 23.4  & -47: 35: 22.2 &   8.37 & 0.16 $\pm$ 0.04 &  No &  No\\
PN  &  120 &              13: 26: 13.0  & -47: 35: 56.3 &   8.64 & 0.22 $\pm$ 0.04 &  No &  No\\
PN  &  143 &              13: 25: 52.5  & -47: 36: 12.2 &   6.88 & 0.19 $\pm$ 0.05 &  No &  No\\
PN  &   46 &              13: 26: 23.8  & -47: 36: 13.5 &   4.16 & 0.36 $\pm$ 0.05 & Yes & Yes\\
PN  &   45 &              13: 26: 39.1  & -47: 36: 33.5 &   3.76 & 0.33 $\pm$ 0.05 & Yes & Yes\\
PN  &   10 &              13: 26: 11.4  & -47: 37: 11.3 &   1.35 & 2.51 $\pm$ 0.13 & Yes & Yes\\
PN  &  142 &              13: 26: 26.6  & -47: 37: 12.7 &   9.22 & 0.17 $\pm$ 0.04 &  No &  No\\
PN  &  134 &              13: 27: 25.0  & -47: 37: 15.1 &  11.10 & 0.21 $\pm$ 0.05 &  No &  No\\
PN  &  113 &              13: 27: 00.3  & -47: 37: 15.2 &   7.51 & 0.15 $\pm$ 0.04 &  No & Yes\\
PN  &   81 &              13: 26: 36.4  & -47: 37: 30.2 &   5.59 & 0.25 $\pm$ 0.05 & Yes &  No\\

\end{tabular}

\end{center}

\end{table*}

\setcounter{table}{1}

\begin{table*}[!th]

\caption{Continued}
\begin{center}

\begin{tabular}{cccccccc}

Instrument & Source &  RA   & Dec   & Position & Rate  &  MOS1 & MOS2 \\
 & ID  & h  m  s  & $\degr$ $\arcmin$ $\arcsec$  & Error (\arcsec)  &  & detection & detection \\

PN  &  128 &              13: 26: 24.1  & -47: 37: 45.7 &   9.76 & 0.23 $\pm$ 0.05 &  No &  No\\
PN  &   19 &              13: 26: 07.4  & -47: 38: 01.3 &   2.31 & 1.15 $\pm$ 0.10 & Yes & Yes\\
PN  &   69 &              13: 27: 33.4  & -47: 38: 35.0 &   5.55 & 0.48 $\pm$ 0.07 & Yes &  No\\
PN  &   58 &              13: 26: 32.4  & -47: 38: 50.7 &   4.94 & 0.38 $\pm$ 0.06 &  No & Yes\\
PN  &   83 &              13: 27: 22.5  & -47: 39: 46.4 &   5.07 & 0.45 $\pm$ 0.07 &  No &  No\\
PN  &   53 &              13: 27: 01.2  & -47: 40: 55.9 &   4.34 & 0.57 $\pm$ 0.07 & Yes & Yes\\
PN  &   66 &              13: 26: 57.9  & -47: 41: 22.4 &   6.22 & 0.54 $\pm$ 0.07 &  No & Yes\\
PN  &  125 &              13: 26: 33.1  & -47: 41: 29.2 &  49.94 & 0.29 $\pm$ 0.06 &  No &  No\\
PN  &  112 &              13: 27: 03.3  & -47: 42: 41.9 &   8.30 & 0.46 $\pm$ 0.07 &  No &  No\\
MOS   &   7 &              13: 25: 52.1  & -47: 19: 09.5 &   0.91 & 2.30 $\pm$ 0.09 & Yes & Yes\\
MOS  &   63 &              13: 25: 40.6  & -47: 22: 08.4 &   4.28 & 0.23 $\pm$ 0.04 & Yes & Yes\\
MOS  &   48 &              13: 25: 34.5  & -47: 25: 58.5 &   3.58 & 0.28 $\pm$ 0.04 & Yes & Yes\\
MOS  &   13 &              13: 26: 20.4  & -47: 30: 03.8 &   1.57 & 0.40 $\pm$ 0.03 & Yes & Yes\\
MOS  &   89 &              13: 26: 25.7  & -47: 39: 04.5 &   4.63 & 0.12 $\pm$ 0.03 & Yes & Yes\\
MOS  &   55 &              13: 26: 10.4  & -47: 39: 10.8 &   3.67 & 0.27 $\pm$ 0.04 & Yes & Yes\\
MOS  &   71 &              13: 26: 10.6  & -47: 40: 28.5 &   5.08 & 0.21 $\pm$ 0.04 & Yes & Yes\\

\noalign{\smallskip\hrule\smallskip}
\end{tabular}

\end{center}

\end{table*}

\begin{table*}[t!]

\caption{The \xmm\ sources correlated with the Chandra and ROSAT sources. The error on the Chandra position is given at the 90\% confidence level, as provided by the {\it wavdetect} task. The ROSAT source numbers are taken from \citet{joh94,ver00}. The EINSTEIN sources are labeled as in \citet{her83}. The source list is declination sorted. \label{gendre_table3}}
\begin{center}

\begin{tabular}{ccccccc}

Source & RA   & Dec   & Chandra  &  Chandra & Position & Previous\\
ID  &   &  & RA   & Dec & error (\arcsec)& ID\\
 & h  m s  & $\degr$  $\arcmin$ $\arcsec$ &     h  m  s  &  $\degr$  $\arcmin$  $\arcsec$ & (RA/Dec) &  \\
\noalign{\smallskip\hrule\smallskip}

  3 &              13: 27: 27.7  & -47: 19: 08.8 &      \nodata     &     \nodata &    \nodata &         4,D \\
  7 &              13: 25: 52.1  & -47: 19: 09.5 &      \nodata     &     \nodata &    \nodata &         3,A \\
 17 &              13: 26: 23.5  & -47: 19: 27.9 &   13: 26: 23.5  & -47: 19: 24.1 & 0.52/0.45 &      \nodata \\
 28 &              13: 26: 46.3  & -47: 19: 47.4 &   13: 26: 46.2  & -47: 19: 45.9 & 0.68/0.39 &          18 \\
 31 &              13: 26: 38.5  & -47: 20: 01.7 &   13: 26: 38.5  & -47: 19: 58.3 & 0.68/0.48 &      \nodata \\
 73 &              13: 26: 35.7  & -47: 20: 57.4 &   13: 26: 35.2  & -47: 20: 52.8 & 1.28/0.64 &      \nodata \\
 68 &              13: 27: 10.2  & -47: 21: 29.2 &   13: 27: 09.9  & -47: 21: 27.7 & 0.78/0.47 &      \nodata \\
 32 &              13: 26: 54.5  & -47: 22: 04.7 &   13: 26: 54.5  & -47: 22: 04.6 & 0.24/0.10 &      \nodata \\
 15 &              13: 26: 41.5  & -47: 22: 16.5 &   13: 26: 41.5  & -47: 22: 16.3 & 0.52/0.31 &      \nodata \\
 39 &              13: 26: 23.2  & -47: 22: 52.7 &   13: 26: 23.1  & -47: 22: 51.2 & 0.73/0.80 &      \nodata \\
 18 &              13: 27: 21.2  & -47: 23: 24.4 &   13: 27: 21.1  & -47: 23: 24.5 & 0.63/0.38 &          19 \\
 41 &              13: 27: 07.9  & -47: 23: 35.3 &   13: 27: 08.0  & -47: 23: 33.6 & 0.65/0.39 &      \nodata \\
 67 &              13: 26: 41.3  & -47: 24: 05.0 &   13: 26: 41.2  & -47: 24: 00.9 & 0.87/0.58 &      \nodata \\
 79 &              13: 26: 12.8  & -47: 24: 16.6 &   13: 26: 12.8  & -47: 24: 13.6 & 0.79/0.59 &      \nodata \\
110 &              13: 27: 12.1  & -47: 24: 22.9 &   13: 27: 12.4  & -47: 24: 24.9 & 1.04/0.78 &      \nodata \\
 94 &              13: 27: 28.5  & -47: 24: 25.2 &   13: 27: 28.4  & -47: 24: 22.2 & 0.89/0.58 &      \nodata \\
126 &              13: 26: 37.0  & -47: 24: 30.5 &   13: 26: 37.3  & -47: 24: 31.3 & 1.07/0.62 &      \nodata \\
 59 &              13: 26: 31.4  & -47: 24: 43.6 &   13: 26: 31.3  & -47: 24: 39.4 & 0.62/0.53 &      \nodata \\
 82 &              13: 27: 06.3  & -47: 25: 37.8 &   13: 27: 06.4  & -47: 25: 39.5 & 0.56/0.69 &      \nodata \\
  6 &              13: 27: 29.4  & -47: 25: 55.4 &   13: 27: 29.3  & -47: 25: 54.8 & 0.28/0.15 &           6 \\
 49 &              13: 26: 28.7  & -47: 26: 33.3 &   13: 26: 28.7  & -47: 26: 27.3 & 0.64/0.23 &      \nodata \\
103 &              13: 26: 45.4  & -47: 26: 53.6 &   13: 26: 45.4  & -47: 26: 52.4 & 1.05/0.65 &      \nodata \\
 44 &              13: 26: 52.6  & -47: 27: 15.9 &   13: 26: 52.9  & -47: 27: 15.7 & 1.17/0.52 &      \nodata \\
 24 &              13: 26: 40.8  & -47: 27: 38.6 &   13: 26: 40.5  & -47: 27: 38.7 & 1.60/0.91 &      \nodata \\
 61 &              13: 26: 52.5  & -47: 27: 39.1 &   13: 26: 52.6  & -47: 27: 38.5 & 1.01/0.50 &      \nodata \\
 76 &              13: 26: 04.2  & -47: 27: 41.6 &   13: 26: 04.4  & -47: 27: 42.1 & 0.65/0.78 &      \nodata \\
 47 &              13: 27: 14.5  & -47: 28: 31.3 &   13: 27: 14.5  & -47: 28: 28.8 & 1.01/0.77 &      \nodata \\
 33 &              13: 25: 52.7  & -47: 28: 55.4 &    \nodata      &     \nodata &    \nodata &           8 \\
 92 &              13: 26: 44.3  & -47: 28: 57.4 &   13: 26: 44.3  & -47: 28: 56.2 & 0.71/0.68 &      \nodata \\
  5 &              13: 26: 53.5  & -47: 29: 01.3 &   13: 26: 53.5  & -47: 29: 00.8 & 0.16/0.07 &        9a,C \\
  4 &              13: 26: 19.7  & -47: 29: 11.5 &   13: 26: 19.7  & -47: 29: 11.2 & 0.15/0.12 &         7,B \\
101 &              13: 26: 37.9  & -47: 29: 11.3 &   13: 26: 38.0  & -47: 29: 10.7 & 1.28/0.79 &      \nodata \\
 51 &              13: 26: 06.2  & -47: 29: 18.7 &   13: 26: 05.9  & -47: 29: 22.3 & 0.70/0.74 &      \nodata \\
  2 &              13: 26: 52.2  & -47: 29: 36.1 &   13: 26: 52.2  & -47: 29: 36.1 & 0.15/0.04 &        9b,C \\
 72 &              13: 26: 34.5  & -47: 29: 57.3 &   13: 26: 34.4  & -47: 29: 55.8 & 0.20/0.18 &      \nodata \\
 13 &              13: 26: 20.4  & -47: 30: 03.8 &   13: 26: 20.4  & -47: 30: 04.0 & 0.19/0.19 &      \nodata \\
137 &              13: 26: 15.5  & -47: 30: 15.1 &   13: 26: 16.7  & -47: 30: 56.3 & 1.42/0.52 &      \nodata \\
 84 &              13: 27: 40.0  & -47: 30: 24.0 &   13: 27: 39.6  & -47: 30: 24.4 & 1.30/0.90 &      \nodata \\
 60 &              13: 26: 34.2  & -47: 30: 34.7 &   13: 26: 34.4  & -47: 30: 34.5 & 0.46/0.33 &      \nodata \\
127 &              13: 27: 04.5  & -47: 30: 38.2 &   13: 27: 04.6  & -47: 30: 37.0 & 0.78/0.68 &      \nodata \\
117 &              13: 26: 23.5  & -47: 30: 49.4 &   13: 26: 23.5  & -47: 30: 44.2 & 0.70/0.59 &      \nodata \\
 20 &              13: 26: 37.4  & -47: 30: 52.9 &   13: 26: 37.4  & -47: 30: 54.5 & 0.38/0.22 &      \nodata \\
 38 &              13: 26: 55.0  & -47: 31: 13.2 &   13: 26: 55.0  & -47: 31: 13.7 & 0.59/0.28 &      \nodata \\
  9 &              13: 26: 48.7  & -47: 31: 26.2 &   13: 26: 48.8  & -47: 31: 26.0 & 0.43/0.20 &          21 \\
 12 &              13: 27: 27.4  & -47: 31: 33.7 &   13: 27: 27.4  & -47: 31: 33.2 & 0.41/0.43 &      \nodata \\
121 &              13: 25: 50.9  & -47: 31: 38.9 &   13: 25: 48.9  & -47: 31: 27.4 & 0.66/0.66 &      \nodata \\
108 &              13: 26: 46.6  & -47: 31: 40.5 &   13: 26: 46.5  & -47: 31: 44.4 & 1.25/0.70 &      \nodata \\
 36 &              13: 26: 50.8  & -47: 31: 47.5 &   13: 26: 51.1  & -47: 31: 45.8 & 0.27/0.28 &      \nodata \\
114 &              13: 27: 14.9  & -47: 31: 50.0 &   13: 27: 14.7  & -47: 31: 49.6 & 1.09/0.46 &      \nodata \\
 74 &              13: 27: 21.7  & -47: 32: 07.3 &   13: 27: 21.5  & -47: 32: 05.9 & 1.20/0.40 &      \nodata \\
 35 &              13: 26: 49.6  & -47: 31: 53.2 &   13: 26: 49.5  & -47: 31: 47.8 & 1.22/0.46 &      \nodata \\
 57 &              13: 26: 49.6  & -47: 32: 13.4 &   13: 26: 49.5  & -47: 32: 13.6 & 0.43/0.15 &      \nodata \\

\end{tabular}
\end{center}

\end{table*}

\setcounter{table}{2}

\begin{table*}[t!]

\caption{Continued}
\begin{center}

\begin{tabular}{ccccccc}

Source & RA   & Dec   &  Chandra  &  Chandra & Position & Previous\\
ID  &   &  &  RA   & Dec & error (\arcsec)& ID\\
 & h  m s  & $\degr$  $\arcmin$ $\arcsec$ &    h  m  s  &  $\degr$  $\arcmin$  $\arcsec$ & (RA/Dec) &  \\
\noalign{\smallskip\hrule\smallskip}

  8 &              13: 26: 25.1  & -47: 32: 28.7 &   13: 26: 25.1  & -47: 32: 28.3 & 0.229/0.20 &          10 \\
 16 &              13: 26: 43.9  & -47: 32: 31.0 &   13: 26: 44.1  & -47: 32: 32.3 & 0.315/0.23 &      \nodata \\
 11 &              13: 27: 11.9  & -47: 32: 41.4 &   13: 27: 11.7  & -47: 32: 41.6 & 0.249/0.22 &      \nodata \\
 21 &              13: 25: 57.2  & -47: 32: 51.6 &   13: 25: 57.2  & -47: 32: 50.8 & 0.373/0.38 &      \nodata \\
  1 &              13: 26: 01.5  & -47: 33: 07.4 &   13: 26: 01.5  & -47: 33: 06.3 & 0.164/0.15 &          11 \\
 34 &              13: 27: 10.0  & -47: 33: 21.8 &   13: 27: 10.0  & -47: 33: 21.8 & 0.568/0.34 &      \nodata \\
 14 &              13: 28: 09.1  & -47: 33: 27.0 &       \nodata     &     \nodata &    \nodata &          12 \\
 25 &              13: 26: 11.5  & -47: 34: 04.2 &   13: 26: 11.5  & -47: 34: 03.4 & 0.401/0.34 &      \nodata \\
 40 &              13: 26: 26.9  & -47: 34: 09.1 &   13: 26: 26.7  & -47: 34: 09.5 & 0.794/0.45 &      \nodata \\
 22 &              13: 26: 13.6  & -47: 34: 43.2 &   13: 26: 13.6  & -47: 34: 41.5 & 0.561/0.44 &      \nodata \\
 30 &              13: 26: 27.4  & -47: 34: 54.9 &   13: 26: 27.6  & -47: 34: 56.9 & 0.361/0.29 &      \nodata \\
 23 &              13: 27: 12.7  & -47: 34: 56.4 &   13: 27: 12.9  & -47: 34: 57.1 & 0.324/0.19 &      \nodata \\
 52 &              13: 26: 59.2  & -47: 34: 59.0 &   13: 26: 59.3  & -47: 34: 58.2 & 0.744/0.48 &      \nodata \\
 45 &              13: 26: 39.1  & -47: 36: 33.5 &   13: 26: 39.2  & -47: 36: 32.3 & 0.727/0.46 &      \nodata \\
 10 &              13: 26: 11.4  & -47: 37: 11.3 &      \nodata      &     \nodata &    \nodata &          13 \\
113 &              13: 27: 00.3  & -47: 37: 15.2 &   13: 27: 00.4  & -47: 37: 15.2 & 0.998/0.64 &      \nodata \\

\noalign{\smallskip\hrule\smallskip}

\end{tabular}
\end{center}

NOTE - We have indicated in this table only the sources correlated with other X-ray observations. 
There are three OGLEGC sources which are associated with X-ray sources (OGLEGC 15 and 22 with two 
Chandra sources, OGLEGC 30 with XMM-Newton source 29). Their positions from \citet{kal96} are :

OGLEGC15 13$^h$ 26$^m$ 47.42$^s$ -47$\degr$ 36\arcmin\ 00.4\arcsec

OGLEGC22 13$^h$ 26$^m$ 08.33$^s$ -47$\degr$ 30\arcmin\ 33.0\arcsec 

OGLEGC30 13$^h$ 26$^m$ 53.33$^s$ -47$\degr$ 18\arcmin\ 22.9\arcsec

\end{table*}

\clearpage

\begin{table}[t!]

\caption{Sources for which a flux variation of a factor two or higher has been detected between the Chandra and \xmm\ observations. The source list is declination sorted, as in Table \ref{gendre_table1} and \ref{gendre_table2}. The unabsorbed Chandra and \xmm\ fluxes are computed from the detected count rates using a 0.6 keV blackbody spectral model and are given in units of  $10^{-15}$ \ergscm.\label{gendre_table4}}
\begin{center}
\begin{tabular}{cccc}

Source & XMM & Chandra & Half mass\\
ID  &  flux  & flux & radius\\
   &         &      & source\\

\noalign{\smallskip\hrule\smallskip}

 73 &  7.1 &  3.3 & No\\
 32 & 12.6 & 67.2 & No\\
 39 & 14.2 &  5.3 & No \\
 18 & 25.1 & 11.2 & No \\
 49 &  4.0 & 15.0 & Yes\\
 24 & 14.7 &  4.8 & Yes\\
 47 &  7.7 &  2.3 & No\\
 13 & 11.0 & 35.6 & No\\
117 &  4.0 &  1.2 & No\\
  9 & 40.0 & 13.7 & Yes\\
108 &  4.3 &  2.0 & Yes\\
  8 & 57.4 & 20.7 & No\\
 23 & 21.1 & 43.3 & No\\

\noalign{\smallskip\hrule\smallskip}

\end{tabular}
\end{center}

\end{table}

\begin{table*}[t!]
\caption{Best fit spectral parameters for the brightest sources of the field of view. We have selected those sources which have more than 100 net counts in EPIC-PN and limited our analysis to the 16 sources within twice the half mass radius. Simple models have been used: Power Law (PL), Blackbody (BB), Thermal Bremsstrahlung (TB). For source 3, which has an optical counterpart \citep{coo95}, we used a two temperature Raymond-Smith model (2T). The model parameters are either the photon index or the temperature in keV. Whenever allowed by the statistics, we left the \nh\ as a free parameter of the fit. Otherwise it was frozen at the optical value ($8.4 \times 10^{20}$ cm$^{-2}$). We used either the $\chi^{\scriptscriptstyle 2}$ or Cash statistics in the fit. The bolometric luminosity is computed between 0.5 and 10 keV and is given in units of $10^{31}$ \ergs. \label{gendre_table5}}
\begin{center}
\begin{tabular}{ccccccc}

Source & Best fit & \nh\  & Model & $\chi^{\scriptscriptstyle 2}_{\scriptscriptstyle \nu}$ & Degree of & Bolometric\\
ID  &  model  & ($10^{20}$ cm$^{-2}$) & parameter &  & freedom & luminosity\\

\noalign{\smallskip\hrule\smallskip}

1 & \bb   & $62^{\scriptscriptstyle +21}_{\scriptscriptstyle -21}$  & $1.2^{\scriptscriptstyle +0.1}_{\scriptscriptstyle -0.1}$   & 1.08 & 55 & 191 $\pm$ 7\\
  & \brem & $167^{\scriptscriptstyle +31}_{\scriptscriptstyle -31}$ & $10.7^{\scriptscriptstyle +11.1}_{\scriptscriptstyle -4.0}$ & 1.00 & 55 & 302 $\pm$ 11 \\
  & \pow  & $183^{\scriptscriptstyle +35}_{\scriptscriptstyle -35}$ & $1.7^{\scriptscriptstyle +0.3}_{\scriptscriptstyle -0.3}$   & 1.03 & 55 & 339 $\pm$ 12\\
  &       &              &                                                              &      &    &\\
2 & \brem & (8.4)        & $23.0^{\scriptscriptstyle +29.9}_{\scriptscriptstyle -10.1}$ & 0.72 & 31 & 67 $\pm$ 3\\
  & \pow  & (8.4)        & $1.4^{\scriptscriptstyle +0.2}_{\scriptscriptstyle -0.2}$    & 0.68 & 31 & 72 $\pm$ 3\\
  &       &              &                                                              &      &    &\\
3 & 2T    &  $30^{\scriptscriptstyle +11}_{\scriptscriptstyle -17}$ &  $0.2^{\scriptscriptstyle +0.1}_{\scriptscriptstyle -0.1}$ &1.08 & 28 & 136 $\pm$ 7 \\
  &       &                                                         &  $0.9^{\scriptscriptstyle +0.1}_{\scriptscriptstyle -0.2}$ &     &    & \\
5 & \brem & (8.4)        & $18.7^{\scriptscriptstyle +65.2}_{\scriptscriptstyle -10.0}$ & 0.78 & 17 & 27 $\pm$ 2 \\
  & \pow  & (8.4)        &  $1.4^{\scriptscriptstyle +0.2}_{\scriptscriptstyle -0.2}$   & 0.80 & 31 & 30 $\pm$ 2\\
  &       &              &                                                              &      &    &\\
6 & \brem & (8.4)        & $4.5^{\scriptscriptstyle +4.7}_{\scriptscriptstyle -1.9}$ & 0.83 & 17 & 27 $\pm$ 3 \\
  & \pow  & (8.4)        &  $1.7^{\scriptscriptstyle +0.3}_{\scriptscriptstyle -0.3}$   & 0.81 & 17 & 32 $\pm$ 3 \\
  &       &              &                                                              &      &    &\\
8 & \brem & (8.4)        & $5.8^{\scriptscriptstyle +7.2}_{\scriptscriptstyle -2.5}$  & 0.98 & 19 & 28 $\pm$ 2 \\
  & \pow  & (8.4)        &  $1.7^{\scriptscriptstyle +0.3}_{\scriptscriptstyle -0.3}$ & 0.81 & 19 & 32 $\pm$ 3 \\
  &       &              &                                                              &      &    &\\
9 & \brem & $21^{\scriptscriptstyle +16}_{\scriptscriptstyle -16}$ & $4.4^{\scriptscriptstyle +6.3}_{\scriptscriptstyle -2.1}$ & 0.58 & 10 & 23 $\pm$ 3 \\
  & \brem & (8.4)        & $8.3^{\scriptscriptstyle +17.0}_{\scriptscriptstyle -3.9}$& 0.79 & 11 & 23 $\pm$ 3 \\
  & \pow  & $31^{\scriptscriptstyle +24}_{\scriptscriptstyle -13}$ & 2.0 $\pm$ 0.6 & 0.51 & 10 & 29 $\pm$ 4 \\
  & \pow  & (8.4)        &  $1.5^{\scriptscriptstyle +0.2}_{\scriptscriptstyle -0.2}$  & 0.96 & 11 & 29 $\pm$ 4 \\
  &       &              &                                                              &      &    &\\
12& \bb   & (8.4)        & $0.3^{\scriptscriptstyle +0.1}_{\scriptscriptstyle -0.1}$ & 1.54 & 9 & 5 $\pm$ 1\\
  & \brem & (8.4)        & $1.2^{\scriptscriptstyle +1.7}_{\scriptscriptstyle -0.6}$ & 1.22 & 9 & 7 $\pm$ 2\\
  & \pow  & (8.4)        & $2.4^{\scriptscriptstyle +0.6}_{\scriptscriptstyle -0.6}$ & 1.10 & 9 & 9 $\pm$ 2 \\
  &       &              &                                                              &      &    &\\
15& \pow  & (8.4)        &  $1.0^{\scriptscriptstyle +0.5}_{\scriptscriptstyle -0.5}$ & 0.42 & 6 & 21 $\pm$ 4\\
  &       &              &                                                              &      &    &\\
18& \brem & (8.4)        & $3.7^{\scriptscriptstyle +11.0}_{\scriptscriptstyle -2.1}$ & 1.12 & 5 & 10 $\pm$ 2\\
  & \pow  & (8.4)        & $1.8^{\scriptscriptstyle +0.5}_{\scriptscriptstyle -0.5}$  & 1.01 & 5 & 12 $\pm$ 2 \\
  &       &              &                                                              &      &    &\\
\noalign{\smallskip\hrule\smallskip}
Source & Best fit & \nh\ & Model & C statistic & Number of & Bolometric\\
ID  &  model  & ($10^{20}$ cm$^{-2}$) & parameter &  & PHA bins & luminosity\\
\noalign{\smallskip\hrule\smallskip}

11& \brem & (8.4)        & $3.2^{\scriptscriptstyle +7.7}_{\scriptscriptstyle -3.2}$ & C 8.29 & 11 & 8 $\pm$ 2 \\
  & \pow  & (8.4)        & $1.9^{\scriptscriptstyle +0.7}_{\scriptscriptstyle -0.7}$ & C 8.96 & 11 & 10 $\pm$ 2 \\
  &       &              &                                                              &      &    &\\
13& \pow  & (8.4)        & $0.9^{\scriptscriptstyle +0.3}_{\scriptscriptstyle -0.3}$ & C 1.16 & 8 & 33 $\pm$ 5\\
  &       &              &                                                              &      &    &\\
20& \bb   & (8.4)        & $0.6^{\scriptscriptstyle +0.2}_{\scriptscriptstyle -0.2}$ & C 16.37 & 11 & 6 $\pm$ 2\\
  & \brem & $52^{\scriptscriptstyle +70}_{\scriptscriptstyle -40}$ & $2.0^{\scriptscriptstyle +4.2}_{\scriptscriptstyle -1.3}$ & C 16.70 & 11 & 9 $\pm$ 2 \\
  &       &              &                                                              &      &    &\\
22& \pow  & (8.4)        & $1.6^{\scriptscriptstyle +0.4}_{\scriptscriptstyle -0.4}$  & C 9.18 & 11 & 12 $\pm$ 2 \\
  &       &              &                                                              &      &    &\\
23& \brem & (8.4)        & $0.7^{\scriptscriptstyle +1.1}_{\scriptscriptstyle -0.4}$ & C 12.58 & 10 & 5 $\pm$ 2 \\
  & \pow  & (8.4)        & $2.8^{\scriptscriptstyle +0.9}_{\scriptscriptstyle -0.9}$ & C 12.77 & 10 & 6 $\pm$ 2 \\
  &       &              &                                                              &      &    &\\
24, high state & \pow & (8.4) & $1.2^{\scriptscriptstyle +0.4}_{\scriptscriptstyle -0.4}$ & C 16.2 & 14 & 18 $\pm$ 2 \\
    low state  & \pow & (8.4) & $1.7^{\scriptscriptstyle +1.4}_{\scriptscriptstyle -1.4}$ & C 10.2 & 11 & 4 $\pm$ 2 \\
  &       &              &                                                              &      &    &\\
25 & \brem & (8.4) 	 & $1.4^{\scriptscriptstyle +5.6}_{\scriptscriptstyle -0.8}$ & C 2.50 & 5 & 6 $\pm$ 2 \\
   & \pow  & (8.4)       & $2.2^{\scriptscriptstyle +0.8}_{\scriptscriptstyle -0.8}$ & C 1.81 & 5 & 8 $\pm$ 2 \\

\noalign{\smallskip\hrule\smallskip}

\end{tabular}
\end{center}

\end{table*}

\begin{table*}[t!]

\caption{Spectral fit parameters of the quiescent neutron star binary candidate source 4, using the spectra of the EPIC-PN and EPIC-MOS cameras. The model used was a pure hydrogen neutron star atmosphere model \citep{pav92,zav96}. Parameters between parenthesis were frozen during the fit. The errors are also given at the 90 \% confidence level. A mass of 1.4M$_{\sun}$ was assumed for the neutron star. The luminosities are given in units of $10^{32}$ \ergs. Parameters obtained by \citet{rut02} are also listed. \label{gendre_table6}}
\begin{center}
\begin{tabular}{ccccccc}

Radius & Temp. & Distance & \nh & $\chi^{\scriptscriptstyle 2}_{\scriptscriptstyle \nu}$ & 0.1-5.0 keV & Reference\\
R$_{\infty}$ (km)  & T$_{eff,\infty}$ (eV) & (kpc) & ($10^{20}$ cm$^{-2}$) &  & luminosity & \\

\noalign{\smallskip\hrule\smallskip}

14.3 $\pm$ 2.1 & $66^{\scriptscriptstyle +4}_{\scriptscriptstyle -5}$ & (5) & (9) & \nodata & $5 \pm 2$ &\citet{rut02} \\

13.6 $\pm$ 0.3 & $67^{\scriptscriptstyle +2}_{\scriptscriptstyle -2}$ & (5.3) & 9.0 $\pm$ 2.5 & 1.00 & 3.2 $\pm$ 0.2 &This work \\

\noalign{\smallskip\hrule\smallskip}

\end{tabular}
\end{center}
\end{table*}

\clearpage

\end{document}